\documentclass[usenatbib]{mn2e}

\newcommand{\etal}{{\it et al.}}
\newcommand{\ie}{{\it i.e.}}
\newcommand{\eg}{{\it e.g.}}
\newcommand{\be}{\begin{equation}}
\newcommand{\ee}{\end{equation}}
\newcommand{\bm}{\begin{displaymath}}
\newcommand{\eem}{\end{displaymath}}

\newcommand{\kmsMpc}{km~s$^{-1}$ Mpc$^{-1}$}

\newcommand{\mnras}{MNRAS}
\newcommand{\apj}{ApJ}
\newcommand{\aj}{AJ}

\newcommand{\aap}{A\&A}
\newcommand{\apjs}{ApJS}
\newcommand{\araa}{ARA\&A}

\usepackage{epsfig}
\begin{document}

\title[Galaxy Zoo: Dust in Spiral Galaxies]{Galaxy Zoo: Dust in Spiral Galaxies \thanks{This publication has been made possible by the participation of more than 160,000 volunteers in the Galaxy Zoo project. Their contributions are individually acknowledged at \texttt{http://www.galaxyzoo.org/Volunteers.aspx}.}}
\author[K.L. Masters \etal]{Karen L. Masters$^1$\thanks{E-mail: karen.masters@port.ac.uk}, Robert Nichol$^1$, Steven Bamford$^2$, Moein Mosleh$^{3,4}$, \newauthor 
Chris J. Lintott$^5$, Dan Andreescu$^6$, Edward M. Edmondson$^1$, William C. Keel$^7$, \newauthor 
Phil Murray$^8$, M. Jordan Raddick$^9$, Kevin Schawinski$^{10}$, An\v{z}e Slosar$^{11}$, \newauthor 
Alexander S. Szalay$^9$, Daniel Thomas$^1$, Jan Vandenberg$^9$\\
 $^1$Institute for Cosmology and Gravitation, University of Portsmouth, Dennis Sciama Building, Burnaby Road, Portsmouth, PO1 3FX, UK \\ 
 $^2$Centre for Astronomy \& Particle Theory, University of Nottingham, University Park, Nottingham, NG7 2RD\\ 
 $^3$Department of Physics and Astronomy, University of Sussex, Brighton, East Sussex, BN1 9QH\\
 $^4$Leiden Observatory, Leiden University, P.O. Box 9513, 2300 RA Leiden,The Netherlands\\
  $^{5}$Astrophysics, University of Oxford, Denys Wilkinson Building,
        Keble Road, Oxford, OX1 3RH, UK\\
  $^{6}$LinkLab, 4506 Graystone Ave., Bronx, NY 10471, USA\\
   $^7$Department of Physics \& Astronomy, 206 Gallalee Hall, 514 University Blvd., University of Alabama, Tuscaloosa, AL 35487-0234, USA\\
  $^{8}$Fingerprint Digital Media, 9 Victoria Close, Newtownards, Co. Down,
        Northern Ireland, BT23 7GY, UK\\
  $^{9}$Department of Physics and Astronomy, The Johns Hopkins University,
        Homewood Campus, Baltimore, MD 21218, USA\\
  $^{10}$Einstein Fellow/Yale Center for Astronomy and Astrophysics, Yale University,
        P.O. Box 208121, New Haven, CT 06520, USA\\
   $^{11}$Berkeley Center for Cosmo. Physics, Lawrence Berkeley National Lab. \& Physics Dept.,
        Univ. of California, Berkeley CA 94720, USA }
\date{Accepted by MNRAS}
\pagerange{1--25} \pubyear{2010}

\label{firstpage}

\maketitle
\begin{abstract}

We investigate the effect of dust on spiral galaxies by measuring the inclination--dependence of optical colours for 24,276 well--resolved SDSS galaxies visually classified via the Galaxy Zoo project. We find clear trends of reddening with inclination which imply a total extinction from face-on to edge-on of 0.7, 0.6, 0.5 and 0.4 magnitudes for the $ugri$ passbands (estimating 0.3 magnitudes of extinction in $z$-band).  We split the sample into ``bulgy" (early--type) and ``disky" (late--type) spirals using the SDSS {\tt fracdeV} (or $f_{DeV}$) parameter and show that the average face-on colour of ``bulgy" spirals is redder than the average edge-on colour of ``disky" spirals. This shows that the observed optical colour of a spiral galaxy is determined almost equally by the spiral type (via the bulge-disk ratio and stellar populations), and reddening due to dust. We find that both luminosity and spiral type affect the total amount of extinction, with disky spirals at $M_r \sim -21.5$ mags having the most reddening -- more than twice as much as both the lowest luminosity and most massive, bulge-dominated spirals. An increase in dust content is well known for more luminous galaxies, but the decrease of the trend for the most luminous has not been observed before and may be related to their lower levels of {\it recent} star formation. We compare our results with the latest dust attentuation models of Tuffs et al. We find that the model reproduces the observed trends reasonably well but overpredicts the amount of $u$-band attenuation in edge-on galaxies. This could be an inadequacy in the Milky Way extinction law (when applied to external galaxies), but more likely indicates the need for a wider range of dust--star geometries. We end by discussing the effects of dust on large galaxy surveys and emphasize that these effects will become important as we push to higher precision measurements of galaxy properties and their clustering. 
\end{abstract}


\section{Introduction}

The clear view we enjoy of the extragalactic sky towards the Galactic poles led to an early assumption that most disks of spiral galaxies are largely transparent. Early work supported this idea \citep[e.g.][]{H58}, and it was not challenged until \citet{DDP89} pointed out that the FIR emission observed by the IRAS satellite could not be explained without absorption and re-emission of optical light by significant amounts of dust. Since then several authors have used the inclination dependence of dust extinction to judge the role of dust on the observed properties of spiral galaxies (eg. \citealt{C92,G94,T98,M03}), and it remains unclear even today if most spiral galaxies are predominantly optically thin (low opacity) or optically thick (high opacity). Studies of overlapping galaxies have shown that the relative geometry of the stars and dust plays an important role, which can vary significantly from galaxy to galaxy. They also show that in a given galaxy there may be both optically thick and thin regions which may or may not correlate with patterns in the stellar density (e.g. \citealt{H09}).

In recent years, several authors have revisited this problem using the sheer size and quality of large galaxy surveys like the Sloan Digital Sky Survey (SDSS; \citealt{Y00}). For example, Alam \& Ryden (2002) first pointed out that the red population of SDSS galaxies (selected via a $u-r$ color cut) is contaminated by dust-reddened edge-on spirals, while \citet{S07} studied the dependence of the luminosity function of 61,506 SDSS spiral galaxies (selected using $f_{\rm DeV}$\footnote{The SDSS measures the {\tt fracdeV} or $f_{\rm DeV}$ parameter, which describes the fraction of the light fit by a de Vaucouleurs profile verses an exponential profile. A pure de Vaucouleurs elliptical should have $f_{\rm DeV}=1$, and a pure exponential disk spiral will have $f_{\rm DeV}=0$.}$\leq 0.5$) on inclination, finding that dust extinction caused about 0.5 magnitudes of dimming in $z$-band and 1.2 mags in $u$-band, consistent with what would be expected for optically thick disks. \citet{UR08} used a more stringent cut of $f_{\rm DeV} \leq 0.1$ (arguing that $f_{\rm DeV} < 0.5$ will result in many early-type interlopers) to select 36,162 late-type spirals. They study trends of $u-r$ colour and $r$-band magnitude in a volume limited subset and find $\sim 1.3$ mags of dimming from face-on to edge-on in $r$-band. 

These inclination effects on the global properties of galaxies have been discussed in most detail by Driver et al. (2007) using data from the  Millenium Galaxy Catalogue \citep[MGC,][]{L03} with bulge-disk decompositions \citep{A06} as well as \citet{M09} using SDSS data. Both of these studies highlight that most measured galactic distributions and relationships, especially from the SDSS, are biased by dust effects which can be up to 2-3 magnitudes in shorter wavelength bands for the most inclined galaxies. Even the original target selection for the SDSS Main Galaxies (Strauss et al. 2002), made in $r$-band could be affected by these issues and this likely leads to subtle incompletenesses in the studies of the large--scale clustering of galaxies (since elliptical galaxies and spiral galaxies have different clustering properties).

We have revisited this issue using a new sample of spiral galaxies selected from the Galaxy Zoo project\footnote{\tt www.galaxyzoo.org}. All previous SDSS studies of the inclination effects have used some measured proxy (color, model fits, concentration, Sersic index) to select their spiral or disk galaxies. As discussed above, there is much debate over the best parameter values to use in selecting such galaxies as well as significant scatter between these different proxies (see Appendix A). In contrast, Galaxy Zoo provides robust visual classifications for over $10^5$ objects thanks to the participation of more than 160,000 volunteers \citep{L08}, and in recent Galaxy Zoo papers we have demonstrated the complexity of relating galaxy colors (and other morphological parameters) to these visual classifications, e.g., \citet{B09,Sk09,redspirals}, and \citet{Sc09} who discuss the interesting sub--populations of  ``red spirals" and ``blue ellipticals"  respectively, found in the Galaxy Zoo data. 

In this paper, we study the trends for $\lambda-z$ (ie. optical colours relative to the $z$-band) reddening as a function of axial ratio for a sample of  well--resolved Galaxy Zoo spiral galaxies. While the total extinction in $z$-band is not expected to be zero (for example \citealt{M03} and \citealt{D08} both show non-zero K-band extinction), we avoid combining SDSS and 2MASS (or UKIDSS, \citealt{L07}) data due to worries about systematic differences in the photometric apertures between the SDSS and these near--Infrared (NIR) surveys, as a function of inclination. In the future, such studies would be improved by including the NIR data and thus giving a ``zero extinction" measurement. 

In Section 2, we describe the data from Galaxy Zoo and SDSS, including galaxy photometry, and axial ratios, and discussed biases introduced by our sample selection. In Section 3, we show the observed change in color as a function of axial ratio, and then in Section 4, compare these trends to the dust attenuation model from \citet{T04}. We discuss the implications of the work and conclude in Section 5.

\section{Data and Sample Selection}
\subsection{Galaxy Zoo Classifications}

The Galaxy Zoo (GZ) project uses an internet tool to allow volunteers from the general public to visually classify galaxies observed by SDSS (see \citealt{L08} for details of the sample selection and initial results). In particular,  Galaxy Zoo participants were asked to say if a galaxy was elliptical, spiral, ``don't know'' or a merger. The spiral classification was then divided into either clockwise or anti-clockwise classes (based on the apparent direction of the spiral arms), or ``edge-on/don't know". In total, each SDSS galaxy received an average of 38 separate classifications, with most of the galaxies having at least 20 independent classifications. Based on these classifications, each object is then assigned a likelihood of being either a spiral or an elliptical (or merger or ``don't know"). Results using these Galaxy Zoo classifications have been presented in a series of recent papers \citep{Land08, S09,B09,Sk09,D09a, D09b,Sc09,C09,redspirals}. 

 The Galaxy Zoo project was initiated by the need for reliable visual morphologies of SDSS galaxies - a sample an order of magnitude larger than any which had previous visual classifications. Discussions of why automated methods for classification were deemed insufficient for many scientific purposes can be found in the introductions of both \citet{L08} and \citet{B09}.  We show, in Appendix A, that visual classification is important to avoid the high levels of incompleteness and contamination present in morphological samples selected using simple structural measurements. We define a sample of spiral galaxies using the Galaxy Zoo classification probabilities, $p_{\rm spiral} > 0.8$. We include a redshift cut of $0.01 < z < 0.09$ to reduce the redshift classification bias discussed in Appendix A of \citet{B09}, and we have also  applied the corrections outlined in \citet{B09}. These cuts provide a sample of 79,935 visually classified spiral galaxies, which to $M_r = -20.5$ is a volume limited sample for face-on spirals
  
\subsection{SDSS Photometry}
Our photometric quantities are taken from the SDSS Data Release 6 (DR6, \citealt{A-M08}). We use Petrosian magnitides for the total flux, and model magnitudes for colours. These are corrected for Galactic extinction using the DIRBE dust maps \citep{dirbe} and have a small k-correction applied using {\sc kcorrectv4\_1\_4} \citep{B03,BR07}. Throughout this paper we use a standard cosmology ($\Omega_M = 0.3$, $\Omega_\lambda=0.7$) with $H_0 = 70$\kmsMpc. 

 As a measure of the size of galaxies, we use the radius enclosing 90\% of the Petrosian flux (in $r$-band), $r_{90}$. We also use the concentration index, $c = r_{90}/r_{50}$ (the ratio between the radius enclosing 90\% of the Petrosian flux and 50\% of the Petrosian flux). We remind the reader that a Petrosian radius is defined to be the radius where the local surface brightness (ie. the mean surface brightness in a small annulus) is equal to a constant fraction of the mean surface brightness within that radius \citep{S02}. 
In the absence of dust, or in a completely opaque disk, the Petrosian radius is independent of inclination (see \citealt{S02}). Dust will cause the light profile of a galaxy to become shallower (as described in \citealt{G94}) causing the Petrosian radius, and $r_{90}$, to move outwards. For example \citet{Moll06} show that in B-band sizes can increase by 10-40\% from face-on to edge-on in the presence of dust. 
 
 We also use the SDSS structural parameter {\tt fracdeV} (or $f_{DeV}$). In SDSS, all galaxy light curves are fit with both a de Vaucouleurs and an exponential profile. Model magnitudes are contructed from an optimal combination of these two fits (the linear combination which best fits the data), and $f_{DeV}$ describes the fraction of the light which is fit by the de Vaucouleurs profile. 
 
\subsection{Inclinations}\label{absection}
We use the observed axial ratio ($a/b$, from SDSS) as a proxy for inclination. The exact correspondence between $a/b$ and inclination will vary from galaxy to galaxy. It will depend on the intrinsic axial ratio, $q$,  of the galaxy (\ie~ that which would be measured for $i=90^\circ$) 
and on any variation from $a/b=1$ which would be observed if the galaxy were face-on.  A reasonable intrinsic axial ratio for spirals varies from $q=0.1$ to $0.2$ (\citealt{UR08} measure $q\sim0.22$), while the ellipticity of the face-on disk is usually assumed to be small (it has been measured at $\epsilon \sim 0.08$, or $b/a=0.92$ by \citealt{UR08}) so is neglected. Inclination can therefore be estimated using
\be
\cos^2 i = \frac{(b/a)^2 - q^2}{1-q^2}. \label{inclination}
\ee

We use the $g$-band isophotal $a/b$ reported by SDSS which are fit to the 25th mag/square arcsecond isophote ($r_{25}$). We pick this axial ratio rather than that calculated from the flux weighted moments of the galaxy as it is likely to be a better estimate of the maximum ellipticity  and therefore of the true inclination of the galaxy (since the moments $a/b$ is flux weighted it is likely to reflect the shape in the brightest parts of the galaxy). We pick $g$-band as a compromise between getting the best $S/N$ (which would lead us to pick the default $r$-band) and the fact that bluer bands better trace the shape of the disks of spirals at these redshifts. 

 Obviously, both the bulge and disk components of a spiral affect it's observed axial ratio. The best way to measure the inclination is through the axial ratio of the disk in a bulge-disk decomposition, but in the absence of that data a value of $q$ in Eqn (1) which increases with $B/T$ provides the best possible estimate.

\begin{figure}
\includegraphics[width=84mm]{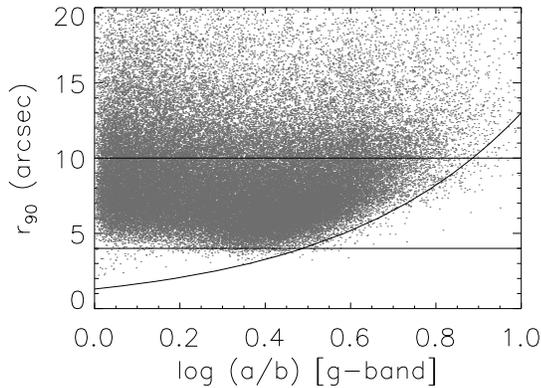}
\caption{The g-band axial ratio versus the radius enclosing 90\% of the $r$-band light for GZ spirals ($p_{\rm spiral } > 0.8$, $0.01\le z \le 0.09$). This shows the effects of seeing which rounds the isophotes of small galaxies. Horizontal lines are plotted at $r_{90}=4\arcsec$ (the apparent minimum of this value for GZ spirals at $0.01 \le z \le 0.09$) and $r_{90}=10\arcsec$, where the effects of seeing appear to disappear almost completely. The curved line shows a simple model in which the minimum possible semi-minor axis is $b=1.3\arcsec$ (an estimate of the 25th percentile best seeing in SDSS - the median is about $1.43\arcsec$).
\label{ab}}
\end{figure}

As discussed on the SDSS Algorithms Page\footnote{www.sdss.org/dr7/algorithms/}. the isophotal axial ratio is not corrected for seeing. We see the impact of this in that the axial ratios of the smallest GZ spirals are on average rounder than the larger spirals. Figure \ref{ab} shows $r_{90}$ against $\log(a/b)$ measured in $g$-band. Horizontal lines are plotted at $r_{90}=4\arcsec$ (the apparent minimum of this value for GZ spirals at $0.01<z<0.09$) and $r_{90}=10\arcsec$, where the effects of seeing appear to disappear. The curved line shows a simple model in which the minimum possible semi-minor axis is $b=1.3\arcsec$ (an estimate of the mean seeing in SDSS), and the major axis is $r_{90}$. This appears to match the observed trend well. It is possible to correct the observed axial ratios for the effects of seeing as was done by \citet{M03} for a sample of 15,224 spirals in 2MASS. However, given the size of our sample, we choose to deal with this issue by simply selecting only large GZ spirals where seeing is not an issue, i.e., $r_{90} > 10\arcsec$. In the rest of this paper, we refer to this selection as our ``well resolved'' GZ spirals ($0.01<z<0.09$) which totals 24,279 galaxies. This selection still admits galaxies over the full redshift and luminosity range above. A Milky Way size galaxy might just make it into the selection at the high redshift limit where $10\arcsec$ corresponds to a physical size of about 17 kpc (it's roughly 2 kpc at the lower redshift limit). However as discussed below in Section 2.5, this strict cut on size will complicate the biases in our sample.

\subsection{Spiral Type or Bulge-Disk Ratio \label{2.4}} 
As has been known since \citet{H22} not all spiral galaxies are the same. Both their physical properties and appearance vary significantly from the ``early type" spirals (Sas with large bulges) to the ``late type" spirals (Sc/Sds with little/no bulge). For a review of this topic see \citet{RH}. Since the star formation histories of different types of spirals differ (with later types having more recent star formation), so do their intrinsic colours. The expected intrinsic axial ratio (how thick they appear when viewed edge-on) is also affected by the presence of a bulge, such that earlier type spirals with large bulges are wider (have smaller $a/b$) when viewed completely edge-on than their late type spirals counterparts.

 Therefore, to fully understand the reddening of spirals with their observed axial ratio we need a way to split galaxies of differing bulge-disk ratios and intrinsic axial ratios. This has been done using bulge-disk decompositions to study the B-band attenuation of 10095 galaxies in the Millennium Galaxy Catalogue \citep{D07}, however SDSS does not provide a bulge-disk decomposition as a standard parameter. We therefore, attempt to use other structural parameters from SDSS to broadly divide the sample into those spirals with a pure disks (\ie~late type spirals) and those with large bulges (\ie~early type spirals). 

\subsubsection{Finding Bulges with Light Profile Information}

 In order to split the sample by ``bulginess'' we use structural parameters provided by SDSS, namely concentration, $c$, and $f_{DeV}$. 
A classic elliptical galaxy is expected to have values of $c\sim 5.5$, and $f_{DeV}=1$ while a pure exponential disk will have $c\sim 2.3$ and $f_{DeV}=0$ \citep{S01}. Based on a sample of $\sim 300$ visually classified galaxies, \citet{S01} recommend that $c=2.6$ be used to divide the population. Furthermore, they argued that a galaxy with $f_{DeV} > 0.5$ is likely to be an elliptical, S0 or Sa galaxy, while those with $f_{DeV} < 0.5$ are late type spirals (Sb or Sc) or irregular. (We explore these cuts with Galaxy Zoo classifications in Appendix A.)

\begin{figure*}
\includegraphics{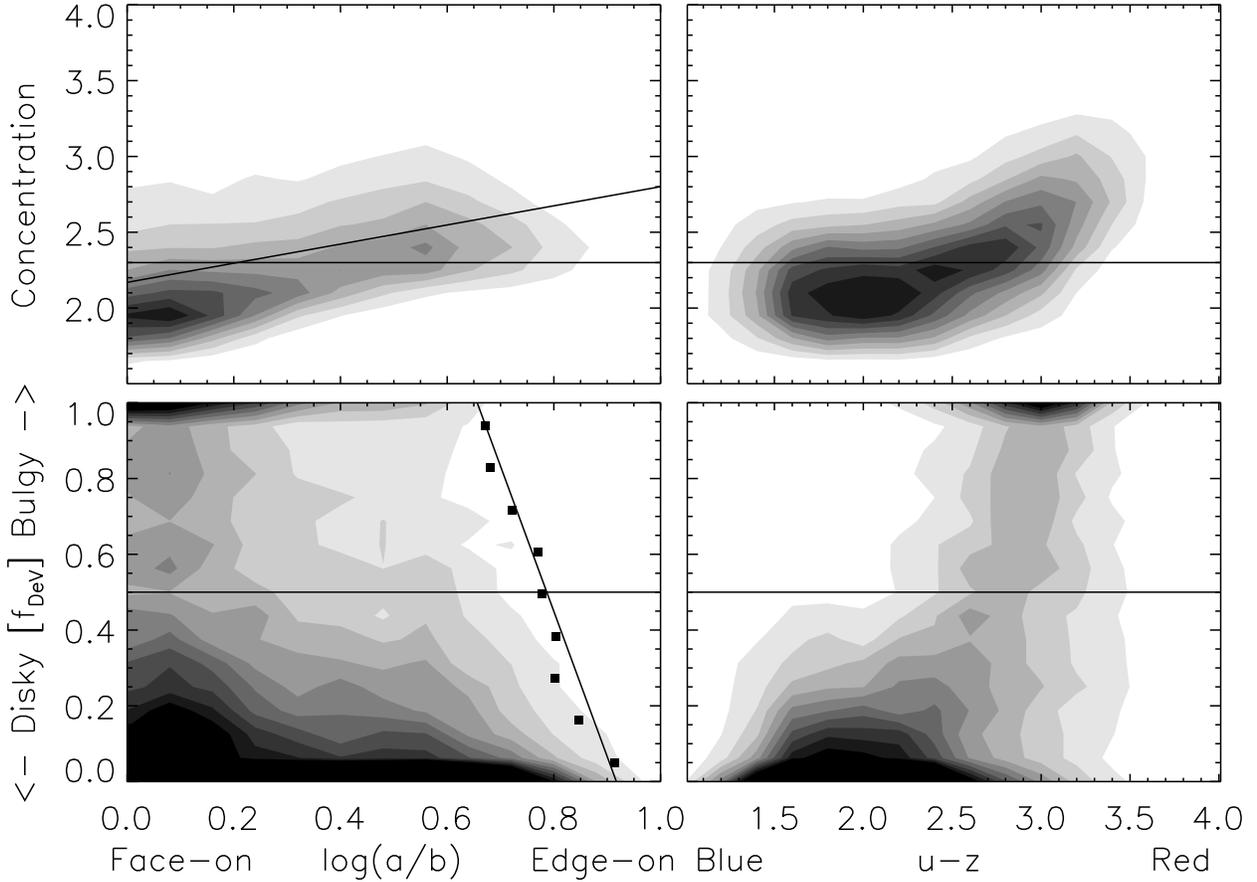}
\caption{Concentration and $f_{DeV}$ for our sample of well resolved GZ spirals as a function of observed axial ratio and $(u-z)$ colour. We plot the best fit to the relation between concentration and axial ratio, and also a fit to our estimate of the relation between the intrinsic axial ratio (ie. the maximum value of $a/b$ possible) with $f_{DeV}$.
\label{bulge}}
\end{figure*}

In Figure \ref{bulge} we plot both concentration and $f_{DeV}$ as a function of axial ratio and $(u-z)$ colour for our sample of ``well resolved'' GZ spirals. It is clear that concentration is not a clean method for spliting our spirals as there is a strong trend such that more inclined spirals are more likely to have high concentrations.  The best linear fit to this trend is $c=2.169(3)+0.632(9)\log(a/b)$ \footnote{We are using a condensed notation for quoting errors on fitted parameters where $2.169(3)$ is equivalent to $2.169\pm0.003$. This is used throughout the paper.}. 
Even the most inclined spirals do not reach the $c=5.5$ expected for a classic elliptical galaxy, however they do reach the lowest concentrations of the observed elliptical range which as discussed in \citet{S01} has a large scatter. 
The mean value of the concentration of Galaxy Zoo spirals reaches $c=2.6$ at $\log (a/b) \sim 0.7$ or $i\sim 80^\circ$ for late type spirals (assumes $q=0.1$), so 50\% of these edge-on spirals will fall into the ``early type" subset if such a divider is used (also see Appendix A).  

On the other hand $f_{DeV}$ appears to be a good candidate to split the spirals into those with large bulges, and those with no bulge. Ironically, \citet{S01} warn about the use of $f_{DeV}$ to split galaxies by type in large, bright galaxies as the model is dominated by light from the central regions, but herein, this is exactly the signal we are looking for, i.e., spirals with large central bulges. We see in Figure \ref{bulge} that ``blue" spirals are found to be only those with small $f_{DeV}$, while ``red" spirals span the full range of $f_{DeV}$ which we interpret as a mix of intrinsically red face-on spirals with large bulges and dust reddened edge-on spirals with no/small bulge. Supporting this interpretation is the fact that there are no galaxies with both large values of $\log(a/b)$ (ie. very thin disks viewed edge-on) and large values of $f_{DeV}$, (i.e., large bulges). 
 
\begin{figure*}
\includegraphics{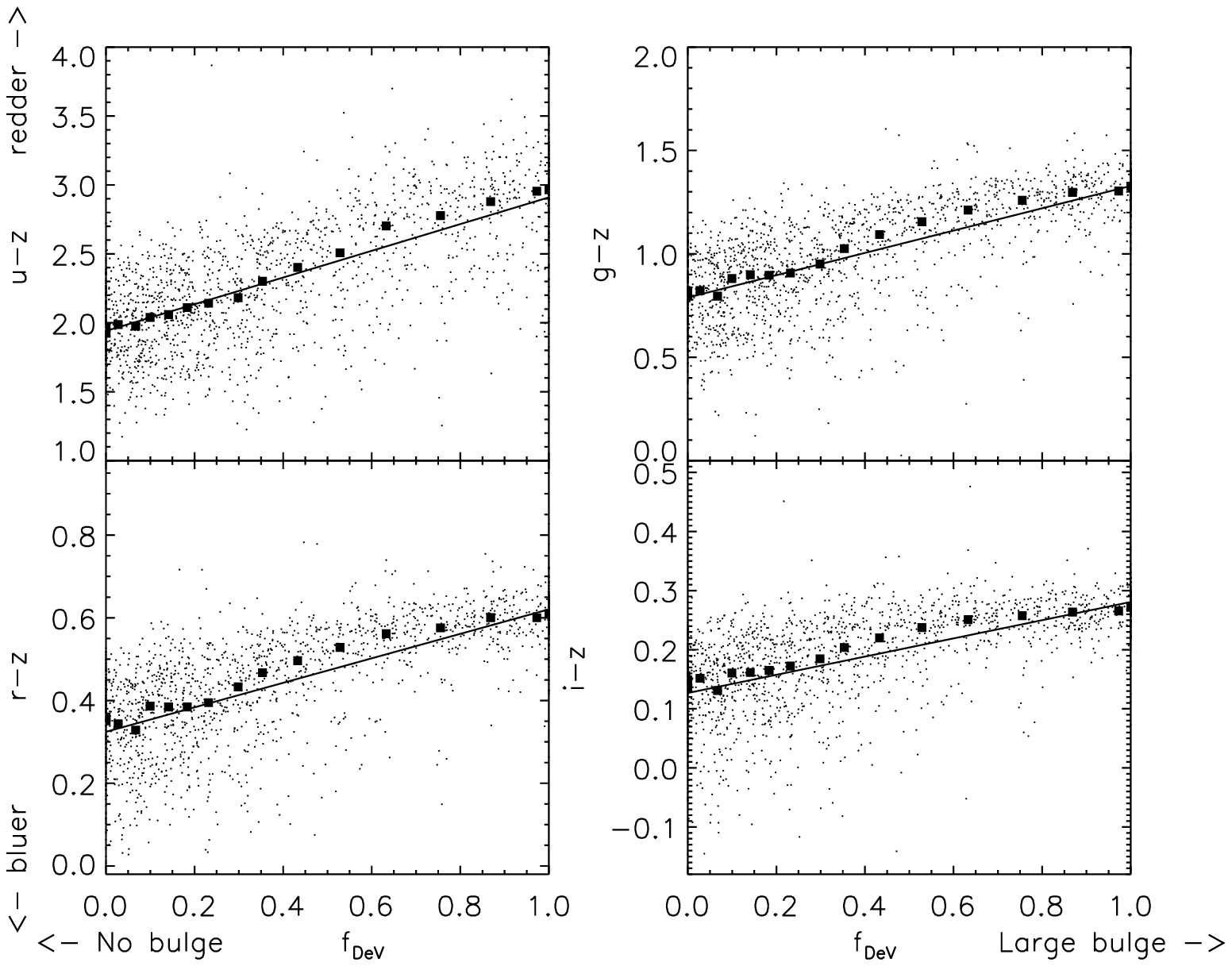}
\caption{The colour of face-on GZ spirals ($\log(a/b)<0.05$) plotted against $f_{DeV}$. The solid line in each panel is a linear fit to the data, and the dots show the medians in 17 bins of 100 galaxies each.
\label{colour_fdeV}}
\end{figure*}
  
As further evidence that the $f_{DeV}$ parameter is able to divide the visually classified GZ spirals by Hubble type (or bulge size), we show in Figure \ref{colour_fdeV} the colours of face-on ($\log(a/b)<0.05$) GZ spirals plotted against their $f_{DeV}$. As expected for early-type spirals (with large bulges), the objects with the largest values of $f_{DeV}$ always have redder (mean) face-on colours than those with small values of $f_{DeV}$, (i.e., spirals with smaller bulges).  At first glance this might seem to contradict the findings of \citet{DF07} who show in a sample of 39 objects that disk galaxies with classical bulges are globally red (irrespective of the bulge size), while only in galaxies with pseudo bulges is there a trend of global colour with $B/T$ (at low values of $B/T$ - since pseudobulges are not found at high $B/T$). This contradiction cannot be clearly tested without identifying the classical and pseudo- bulges in our sample face-on spirals, however looking at Figure \ref{colour_fdeV} in more detail we argue that it may still show the expected trend. There is a larger spread of colours of our face-on spirals with small values of $f_{DeV}$ than is seen in those with large values of $f_{DeV}$ - consistent with the idea of a mix of redder galaxies with classical bulges and bluer galaxies with pseudo-bulges at low $f_{DeV}$ and only redder galaxies with classical bulges at high values of $f_{DeV}$.

We note that our data shows that using a strict cut on $f_{DeV}$ (as done by \citealt{S07} and \citealt{UR08}) will miss a large fraction of the spiral population.  It also shows that an early-type sample selected using a minimum value of $f_{DeV}$ will have significant contamination from visually classified spirals with large values of $f_{DeV}$. We discuss this further in Appendix A.

In Figure \ref{fdeV_fbulge}, we show the match between our well--resolved GZ spirals and the Millennium Galaxy Catalog (MGC, \citealt{L03}), finding an overlap of 109 galaxies (the MGC is much deeper than SDSS but over a smaller area). We use this match to find a linear relation between the $f_{DeV}$ parameter from SDSS and the bulge fraction as measured by the MGC. This relation is shown in Figure \ref{fdeV_fbulge} and is given by $f_{DeV} = 0.11(3) + 1.7(2) (B/T)_{\rm MGC}$
 
\begin{figure}
\includegraphics[width=84mm]{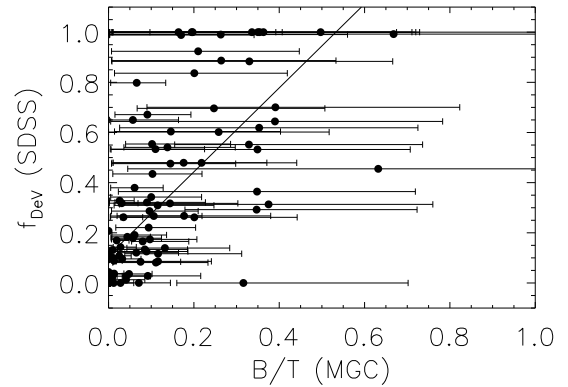}
\caption{SDSS $f_{DeV}$ parameter against the MGC bulge-fraction for the 109 well--resolved GZ spirals found in the MGC. The solid line shows the best fit value of $f_{DeV} = 0.11(3) + 1.7(2)B/T$.
\label{fdeV_fbulge}}
\end{figure}

\begin{figure*}
\includegraphics[height=170mm,angle=-90]{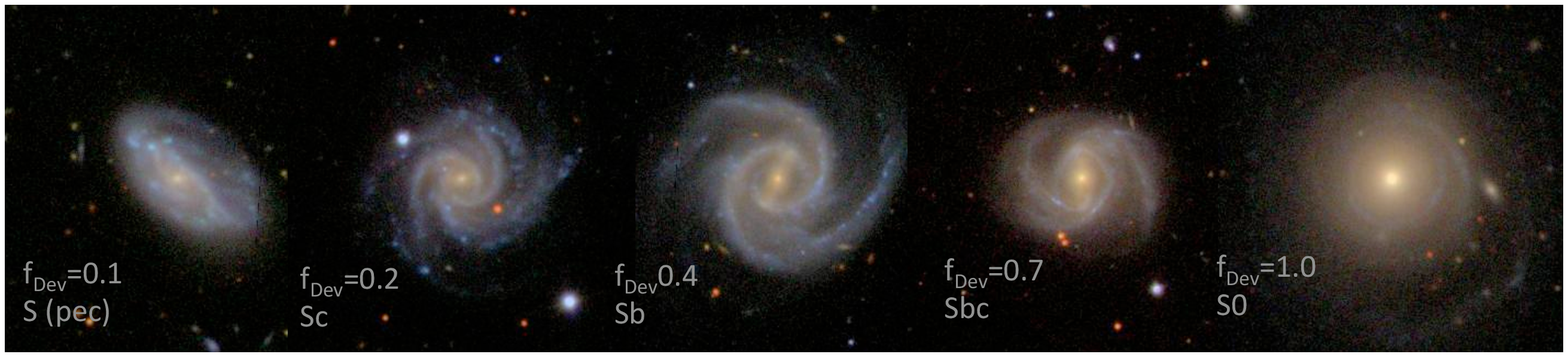}
\caption{Examples of face-on GZ spiral galaxies ordered by $f_{DeV}$
\label{fdeV_images}}
\end{figure*}

 Finally, Figure \ref{fdeV_images} shows example images of face-on GZ spirals of similar angular size, ordered by $f_{DeV}$.

\subsubsection{Intrinsic Axial Ratios}
 
 In order to obtain the inclination of a galaxy from its observed axial ratio, an estimate of its intrinsic axial ratio (ie. the axial ratio which would be observed if it were completely edge-on) is needed. This is observed to vary along the Hubble sequence, with earlier type spirals with large bulges appearing thicker when viewed edge on than later type spirals. \citet{HG84} tabulated the measured values of $q$ for different Hubble types (based on work from an unpublished preprint by Lewis 1980), while \citet{SdV86} give the expected bulge-total luminosity ratios for different Hubble Types. We provide in Table \ref{qbulge} a compilation of these two results in which we also include the approximate value of $f_{DeV}$ we find for these values of $B/D$ in our well resolved GZ spirals.

Here, we use $f_{DeV}$ to predict the intrinsic axial ratio of spiral galaxies. We plot in the bottom left--hand panel of Figure \ref{bulge} an estimate of the maximum axial ratio as a function of $f_{DeV}$  for which we use $\log(a/b)_{\rm max} \equiv \langle \log(a/b) \rangle + 2\sigma_{\log(a/b)}$. We will use a fit to $q = (b/a)_{\rm max}$ from these values as an estimate of the trend of the intrinsic axial ratio of a galaxy with $f_{DeV}$, which we find to be $q = 0.12 + 0.10 f_{DeV}$, consistent with the range of value of $q$ observed in edge-on galaxies in Table \ref{qbulge}. 

\begin{table}
\caption{Bulge-Disk Ratio and Intrinsic Axial Ratio as a Function of Hubble Type. Also included is an estimate of the mean value of $f_{DeV}$ for each type.}
\label{qbulge}
\begin{tabular}{lccccc}
\hline
Type  & T  &  $B/T$ &  $B/D$ &  $q$ & $f_{DeV}$ \\
\hline
S0- & -3 & 0.6 & 1.5 & 0.23 & 1.0 \\
S0  & -2 & 0.57 & 1.3 & 0.23 & 1.0 \\
S0+ & -1 & 0.53 & 1.1 & 0.23 & 1.0 \\
S0/a & 0 & 0.48 & 0.92 & 0.23 & 0.9\\
Sa & 1 & 0.41 & 0.69 & 0.23 & 0.8\\
Sab & 2 & 0.32 & 0.47 & 0.23 & 0.6\\
Sb & 3 & 0.24 & 0.32 & 0.23 & 0.5\\
Sbc & 4 & 0.16 & 0.19 & 0.2 & 0.4 \\
Sc & 5 & 0.094 & 0.10 & 0.175 & 0.3 \\
Scd & 6 & 0.049 & 0.05 & 0.14 & 0.2 \\
Sd & 7 & 0.022& 0.022 & 0.103 & 0.1\\
\hline
\end{tabular}
\end{table}

\subsection{Sample Bias}

Studies of the trends of observed quantities with inclination are hampered by sample bias. Most observed quantities depend in some way on inclination, therefore cuts on observed quantities produce samples biased by inclination. The intercorrelations between properties of galaxies (for example size-magnitude and colour-magnitude relations) also add complications. An unbiased sample for the study of inclination trends is probably not possible, but an appreciation of the effect of any bias in the sample on the results will aid in the interpretation.  

 While the majority of face-on S0s will not make it into the GZ spiral cut, even expert classifiers are more likely to classify S0s as spirals if they are observed edge-on. Therefore we expect the edge-on GZ spirals may have a larger proportion of S0 galaxies than the face-on. A comparison of expert classifications from \citet{F07} and Galaxy Zoo classifications \citep{L08,B09} shows that S0s contribute less than 3\% to the GZ spirals, however if \citet{F07} also tend to classify edge-on S0s as spirals this comparison is not useful. The possible ``contamination" of S0s into the edge-on GZ spirals should be remembered when interpreting our results - since S0s are usually brighter and redder than spiral galaxies this could cause attenuation to be underestimated while reddening measures could be overestimated.

 The sample selected here, based on the SDSS Main Galaxy Sample has an implicit magnitude limit in the $r$-band of $m_r=17.77$. Our redshift limit of $0.01 < z < 0.09$ therefore makes this a volume limited§ sample for face-on spirals to $M_r = -20.5$, while dimmer spirals are only observed in the near parts of the sample. Because of the dimming effects of dust, any magnitude limited sample will miss intrinsically brighter objects at large inclinations. 
 
  We also apply a size cut to our sample. (As discussed above, because of the effects of seeing, we cut at an observed Petrosian 90\% flux semi-major axis (in $r$-band) of $r_{90}=10\arcsec$.)  The presence of dust in a galaxy disk shallows the light profile and moves $r_{90}$ outwards as a galaxy becomes more inclined. This size cut then allows intrinsically smaller objects into the sample at larger inclinations. There is an observed correlation between size and magnitude (most small galaxies are also dimmer) however there is also scatter in that relation, so the net effect here is to have a sample of inclined galaxies which is intrinsically brighter and smaller (\ie~ more compact) than the comparison face-on galaxies. 

\begin{figure}
\includegraphics[width=84mm]{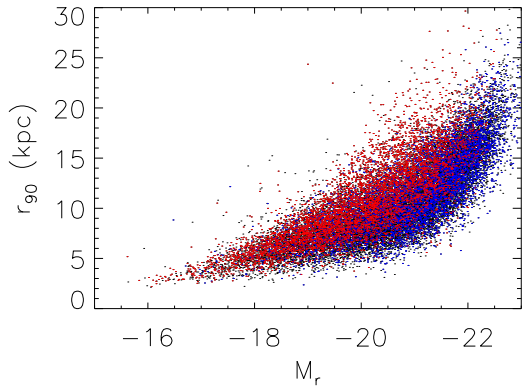}
\caption{Observed physical size versus absolute $r$-band magnitude (corrected for Galactic dust and k-corrections, but not internal extinction) for our sample of well resolved GZ spirals. The most face-on quartile are hi-lighted in blue, while the most edge-on quartile are hi-lighted in red.
\label{size-mag}}
\end{figure}

\begin{figure}
\includegraphics[width=84mm]{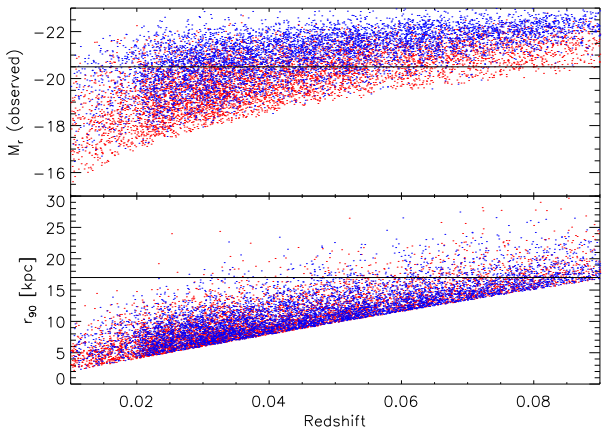}
\caption{Absolute $r$-band magnitude (upper panel) and physical size (lower panel) of the sample of well resolved GZ spirals discussed in this paper. Only the most face-on (blue points) and most inclined (red points) quartiles are shown to illustrate the differences in the samples at the two extremes of inclination. 
\label{sample}}
\end{figure}

 We show in Figure \ref{size-mag} the observed size-magnitude relation for our sample of well resolved GZ spirals. As is well know, brighter galaxies are physically larger. As expected, the most inclined galaxies in our sample have observed sizes which are larger than the more face-on at a given observed (k-corrected and Galactic extinction corrected) magnitude. We expect that intrinsically these two populations should follow the same relation, so this illustrates the effect of inclination on our sample selection - it dims the observed magnitudes and increases the radii outwards at the same time. To make these two relations align, the magnitudes of the edge-on spirals must be shifted by $\sim$0.5--1.0 magnitude (ignoring any change in size), quite consistent with the results on total $r$-band extinction found here (see Section 3) and other studies \citep[e.g.][]{UR08}. Figure \ref{sample} shows the result of the two cuts (SDSS Main Galaxy Sample magnitude limit, and our applied radius limit) on the absolute magnitude and linear sizes of the sample as a function of redshift. Only the most face-on and edge-on quartiles of the sample are shown to illustrate the differences. At all redshifts the observed magnitudes of the inclined galaxies are dimmer than those of the face-on galaxies (which shows the combined effect of the magnitude cut excluding brighter edge-on galaxies and our size cut which allows in inclined galaxies at observed dimmer magnitudes if they have increased observed radii). The intrinsic magnitudes of the inclined galaxies must be brighter than their observed magnitudes, so the distribution of intrinsic magnitudes across the inclination range of our sample may be more similar. 
 
\begin{figure}
\includegraphics[width=84mm]{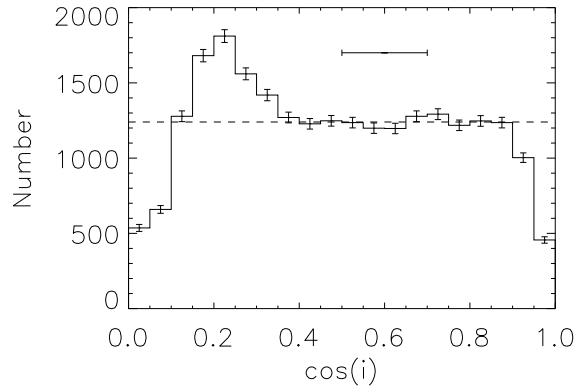}
\caption{Estimate of the $\cos i$ distribution of the full sample. Poisson counting errors are shown, as is the expected flat distribution (dashed line). The estimated median error on the value of $\cos i$ is shown by the horizontal bar. 
\label{cosi}}
\end{figure}

Converting observed axial ratios to inclinations depends on assumptions about the intrinsic axial ratio of the galaxy (as discussed in Section 2.4.2 above). In order to provide empirical corrections, and compare observed data to models, we avoid making this conversion (which adds an additional source of error), although we provide a prescription for estimating the intrinsic axial ratio from $f_{DeV}$ in Section 2.4.2 which we will apply here. The $\cos i$ distribution of the sample is a good sanity check of the level of sample bias. This distribution should be flat (\ie~ random orientations) if the sample is completely unbiased. We show this distribution in Figure \ref{cosi}. There is an excess of objects at $i \sim 78^\circ$, and a deficit at $i < 25^\circ$ and $i>84^\circ$. However we estimate a typical error of $\pm 0.1$ in the estimate of $\cos i$ (coming from an error of $\pm 0.1$ in the measurement of $b/a$ and $\pm 0.05$ in the estimate of $q$) and argue the distribution is reasonably flat within this error. For a very conservative interpretation of the results shown below the region $25^\circ < i < 70 ^\circ$ can be considered completely unbiased (roughly $0.04<\log(a/b)<0.45$).

 That this sample ends up being relatively unbiased (as shown by Figure 8) can be explained if either the biases introduced by the two selection effects (SDSS magnitude limit and our size cut) cancel out precisely, or if one of the selections dominates and this selection is relatively unbiased. The first explanation, while possible, seems unlikely. In fact, Figures 6 \& 7 seem to show that the size selection is the dominant selection effect (\ie~ most objects not making the SDSS magnitude cut would not have made the size cut anyway). This is most easily argued from Figure 6, since a shift in $r$-band magnitude of $\sim0.5$-1.0 aligns the size-magnitude relations for edge-on and face-on spirals with no need for size shifts, and this is in agreement with amount of total attenuation estimated in $r$-band by us and other authors (see Section 3, and e.g. \citealt{UR08}). This then suggests that inclination has little effect on the observed sizes of spiral galaxies -- at least at the radius traced by $r_{90}$. This is possible, if the galaxies are mostly opaque within this radius in the optical bands. \citet{M03} using 2MASS data, asked if spiral galaxies are transparent in NIR bands at 2-3 scale lengths, and conclude this is unlikely in the J, and H-bands (but consistent with the data at K-band).  The radius, $r_{90}$ (enclosing 90\% of the Petrosian flux) is at $\sim4$ scale lengths for a pure exponential \citep[Figure 1]{S02}, but it still does not seem inconsistent that spirals would still be relatively opaque in the shorter optical bands at this radius. It seems therefore that the size cut we apply results in a relatively unbiased sample.
 
\section{Effect of Viewing Angle on the Colours of Spiral Galaxies \label{3}}

 We observe disk galaxies at a variety of inclinations ranging from completely face-on (with axial ratios $\sim1$, so $\log(a/b)=0$) to completely edge on (with large axial ratios, \eg~ $a/b=10$ corresponding to $\log(a/b)=1.0$). The universe is assumed to be isotropic and homogeneous, so we expect that the intrinsic properties of galaxies should not vary with viewing angle. Any such variation in an unbiased sample can therefore be largely interpreted as the effect of an increased path length through the disk of the galaxy (with complications due to the different distributions of stars and dust in a galaxy, which may be averaged out in a large sample). The increased path length in the presence of dust will result in reddening, and dimming, of the emerging light.
  
 In this section, we plot the variation of colour with axial ratio (as a proxy for inclination) in our sample of GZ spirals. We use a simple linear parameterization with $\log(a/b)$ to quantify the effect of inclination on the observed magnitudes and colours of the galaxies. This is the parameterization that has been historically used in studies of inclination dependent effects, presumably arising from the fact that for a totally opaque, pure exponential disk the total magnitude will change by $2.5 \log(a/b)$. We write,
\be
X_{\rm true} = X_{\rm obs} - \gamma_X \log(a/b).
\ee
where $X$ describes the band in question ($X=ugriz$ for SDSS bands), and $\gamma_X$ for an individual galaxy might reasonably be expected to depend on things like the galaxy mass and/or luminosity, its intrinsic colour, its metallicity, its Hubble type, etc.

\begin{figure*}
\includegraphics{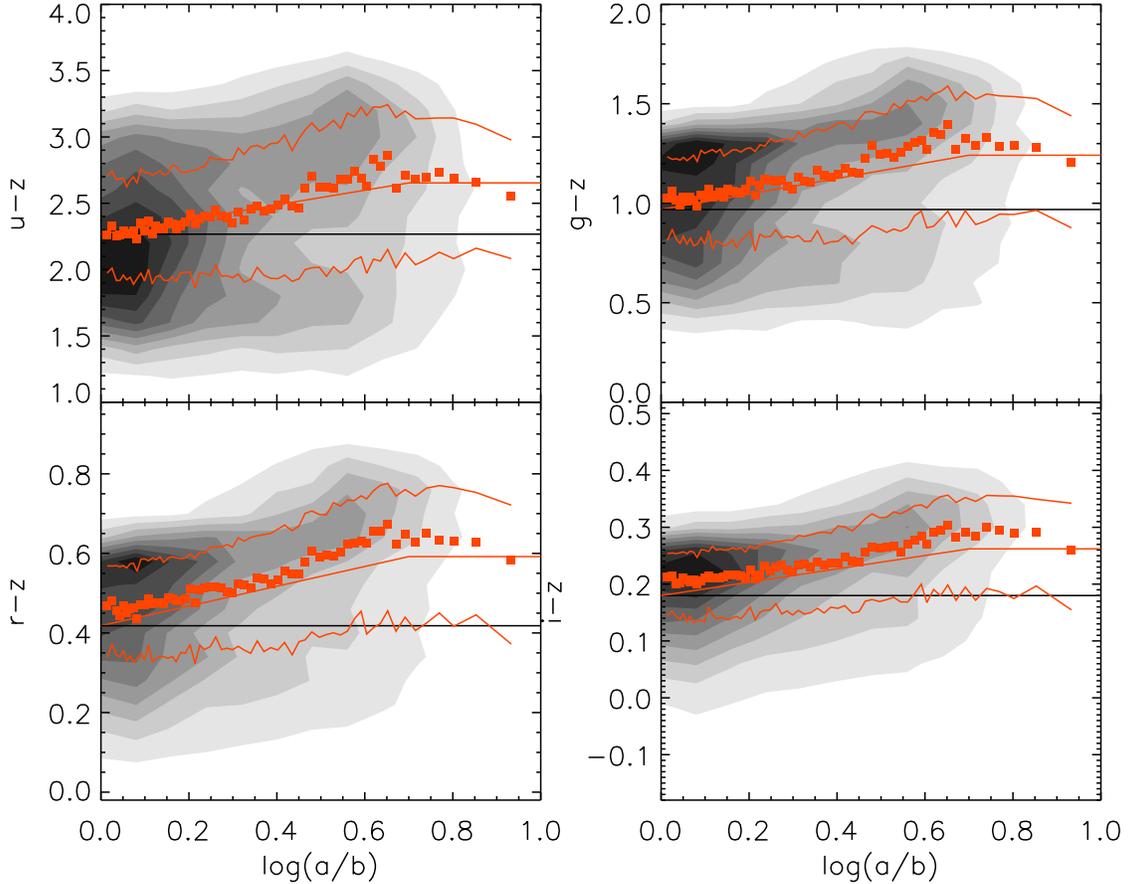}
\caption{Shown here are the trends of $u-z$, $g-z$, $r-z$ and $i-z$ colours as a function of axial ratio for our sample of  well--resolved GZ spirals. Greyscale contours are linear in the density of all galaxies. The median in bins of 400 galaxies are overlaid, as are the 25th and 75th percentiles. The best fit linear relation (to $\log(a/b)<0.7$) is shown as the straight line; for $\log(a/b)>0.7$ we just show a constant extension of this fit. Note that the y-axis range, set by the overall dispersion in colour differs dramatically between $(u-z)$ where is is 3 magnitudes, to $(i-z)$ where is is only 0.7 magnitudes. 
\label{colour}}
\end{figure*}

In Figure \ref{colour}, we show the effect of inclination on the colour of our spiral galaxies. As expected we find a strong trend such that more elongated (or inclined) galaxies are reddened. We fit our linear parameterization, as discussed above, in the range of $\log(a/b) <0.7$ and find;
\begin{eqnarray}
u-z & = & 2.267(7) + 0.55(2)\log(a/b),  \nonumber \\
g-z & = & 0.969(4) + 0.39(1)\log(a/b),  \nonumber \\
r-z & = & 0.418(3) + 0.25(1) \log(a/b),  \nonumber \\
i-z & = & 0.180(2) + 0.12(1) \log(a/b).
\end{eqnarray}

The trends of other colours can be found using linear combinations of these relationships. Note that the errors here are statistical - the systematic error due to the possible sample biases discussed in Section 2.5 is likely larger. 

The scale of the $y$--axis in the different panels of Figure \ref{colour} has been set by the overall dispersion in the colours of our galaxies, and varies from 3 magnitudes in $u-z$ to only 0.7 magnitudes in $i-z$. The trend of the reddening due to dust, while clearly present in all the observed colours shown here, is on average smaller than the overall dispersion seen in this sample of visually selected spirals (which is a mixture of all Hubble types). It appears that some galaxy colours can be massively affected by dust, while others could be affected very little at all. This plot illustrates however that the stellar populations (mostly through the size of bulge, or old stellar component), are just as important in explaining the colors of these well--resolved spiral galaxies as the amount of dust in their disks. Dust should be considered just one of many factors explaining the colours of spirals - however its systematic reddening of spiral colours with inclination means it must be dealt with if inclined systems are to be included in studies of galaxy evolution.

At $\log(a/b) > 0.7$, the colour trends above are suppressed by a downturn in the mean colours for the most elongated spirals.  We interpret this effect as being due to the fact that only the latest, and bluest, of the spiral types (\ie~ Scs and Sds with no bulge component) can have such elongated shapes. As discussed below, an early--type spiral will reach $i=90^\circ$ at $\log(a/b) \sim 0.7$ so cannot have any large value of $a/b$. This illustrates the difficulties of using axial ratio as a proxy for inclination in a sample which includes a wide range of spiral galaxy types. The average colours of the most inclined galaxies may also be impacted by our $r_{90}$ selection which allows in intrinsically smaller galaxies at high inclinations, which on average will also be less luminous and possibly bluer (through the colour-luminosity relation). It could also be observed if the disks become totally opaque at these high inclinations and the light is therefore dominated by bluer emission from the near side disk. 

Our observed trends of $\lambda-z$ with $\log(a/b)$ can provide a lower limit to the average value of $\gamma_X$ for spiral galaxies (under the assumption that the internal extinction in $z$-band is negligible). The SDSS $z$-band should have similar (or slightly more) internal extinction than the J-band, which was measured as $\gamma_J \sim 0.5$ for a sample of visually classified spirals detected in 2MASS by \citet{M03}, thus suggesting $\gamma_z > 0.5$. \footnote{\citet{M09} argue for much lower extinction in J-band ($\max(\gamma_J) = 0.17$), but do quote $\max(\gamma_z) = 0.53$, so assuming $\gamma_z > 0.5$ is not inconsistent with that work.} Using this argument, we use the above SDSS colour trends to show that, on average (neglecting dependences on galaxy mass, luminosity, colour, metallicity and type), $\gamma_u > 1.0$, $\gamma_g > 0.9$, $\gamma_r > 0.8$, $\gamma_i > 0.6$ which we suggest should be used to $\log(a/b)\sim0.7$ after which the extinction appears to be on average constant with axial ratio. This means that a lower limit on the total extinction of spiral galaxies from face-on to edge-on, is 0.7, 0.6, 0.5 and 0.4 magnitudes in $ugri$ bands respectively (based on an assumption of at least $\sim$ 0.3 magnitudes of extinction in $z$-band from face-on to edge-on).

 We advise against using uncorrected axial ratios as a proxy for inclination in galaxies smaller than $r_{90}=10\arcsec$ due to the effects of seeing which will make the most inclined galaxies appear rounder than they should be (as shown in Figure 1). This effect will compress the trend of reddening with inclination making it appear larger than it really is. However we recognize that a correction to the mean colours to the face-on value may still be of use in statistical studies of galaxy evolution so provide below trends which are for ``small GZ spirals'' (meaning those with $r_{90} < 10\arcsec$ only, or  55,653 galaxies in total):
\begin{eqnarray}
u-z & = & 1.980(4) + 1.221(13)\log(a/b) \nonumber \\ 
g-z & = & 0.863(3) + 0.692(7)\log(a/b) \nonumber \\
r-z & = & 0.380(2) + 0.388(4) \log(a/b) \nonumber \\
i-z & = & 0.188(1) + 0.165(2) \log(a/b).  
\end{eqnarray}

\subsection{Dependence on Bulge-Disk Ratio or Spiral Type}

 In Figure \ref{colour}, there is evidence for a possible bimodality in the trend of the colours with axial ratio. As we will see below, when discussing the photometric model of \citet{T04}, the bulge-disk ratio (or spiral type) affects the model attenuation curves. We also know that the average colours of spiral galaxies becomes bluer as they progress along the Hubble sequence (from early to late spirals) as their bulges become smaller. So an explanation for the apparent split in the colour-$\log(a/b)$ trend may be the bulge-disk ratio of the spirals (or perhaps bulge type - see \citealt{DF07}). 
 
\begin{figure*}
\includegraphics{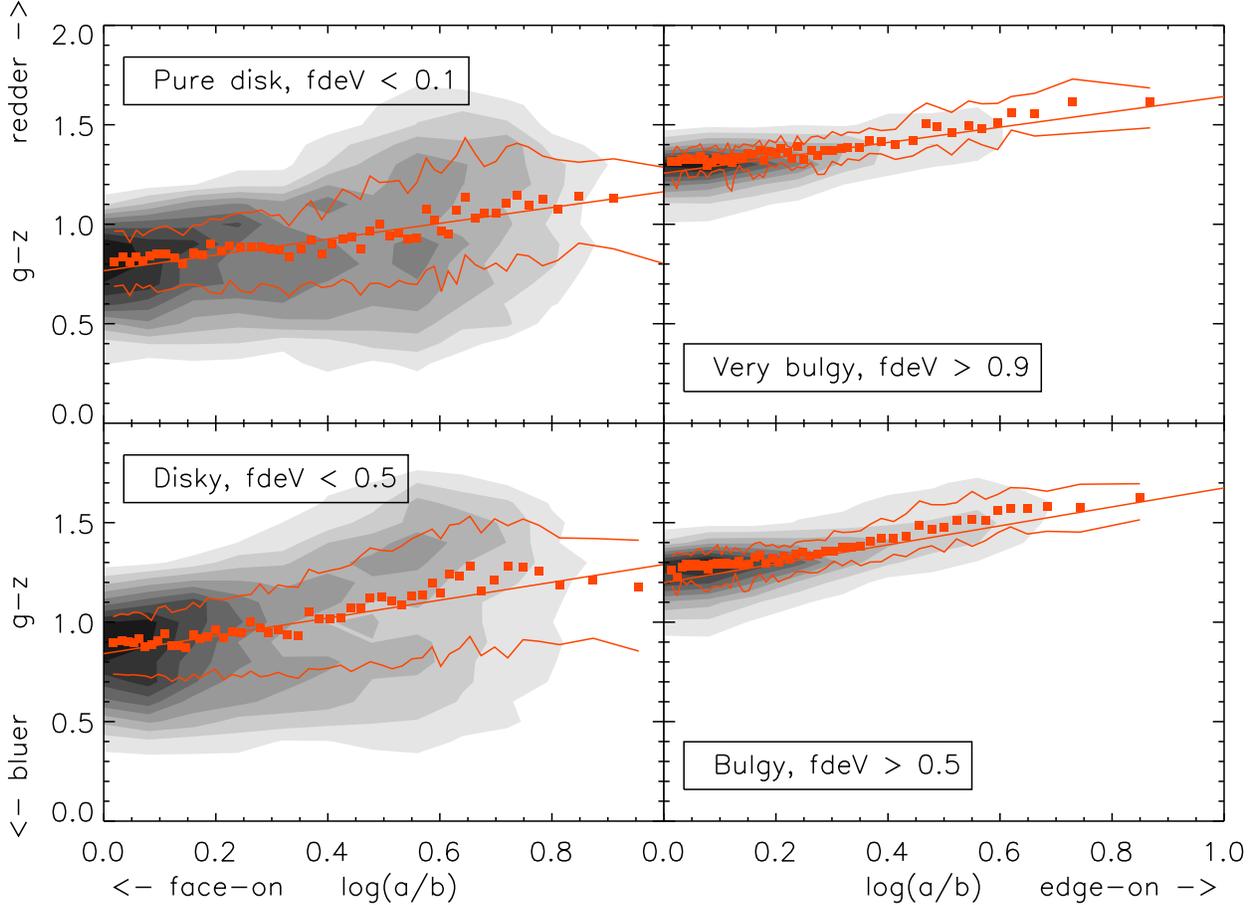}
\caption{The trend of $g-z$ colour for GZ spirals separated using the $f_{DeV}$ parameter as indicated in the panels. As in Figure \ref{colour}, the greyscale contours represent the locus of galaxies, while overlaid are the median and interquartile range of data binned in $\log(a/b)$ (bin sizes are (1) $f_{DeV}<0.1$; 163, (2) $f_{DeV} < 0.5$; 352, (3) $f_{DeV} > 0.5$; 134, (4) $f_{DeV} > 0.9$; 50 - designed to make $\sim 50$ bins for each subset)
\label{c1}}
\end{figure*}

 As discussed in Section 2.4 above, we can use $f_{DeV}$ to separate GZ spirals into subsamples based on the size of their bulge (or give a crude spiral type classifications). Obviously it would be better to use full bulge-disk decompositions to study these effects which would be a natural extension of this work. In Figure \ref{c1}, we show the trends of $g-z$ colour with axial ratio in four bins of $f_{DeV}$ (see Figures \ref{c2}-\ref{c4} in Appendix B for other colour trends). In Table \ref{ctable}, we present the best fit parameters to these relations. While for the full sample, we only fit the range $\log(a/b)<0.7$, for these subsamples (split by  $f_{DeV}$) we fit the full range of $\log(a/b)$ as  the flattening in the median colours seen in Figure \ref{colour} is less pronounced here - perhaps because of the split into rough spiral types.

\begin{table*}
\caption{Best Fit linear Trends to $(\lambda - z) = C_\circ + \gamma \log(a/b)$. $\Delta_0$ is the total reddening from face-on to edge-on.}
\label{ctable}
\begin{tabular}{lccccccccccccc}
\hline 
Sample & $N_{\rm gal}$ & \multicolumn{3}{c}{$(u-z)$} &  \multicolumn{3}{c}{$(g-z)$} &  \multicolumn{3}{c}{$(r-z)$} &  \multicolumn{3}{c}{$(i-z)$} \\
  &   &  $C_\circ$ & $\gamma$& $\Delta_0$ & $C_\circ$ & $\gamma$ & $\Delta_0$ & $C_\circ$ & $\gamma$  & $\Delta_0$ & $C_\circ$ & $\gamma$& $\Delta_0$  \\
\hline
All big spirals &  24276 & 2.267(7) & 0.55(2) & 0.39  & 0.969(4)  & 0.39(1) &  0.27 & 0.418(3)  & 0.25(1) & 0.18  & 0.180(2) & 0.12(1) & 0.08 \\
``Very bulgy''  &  2519 & 2.81(1)   & 0.62(4) & 0.43 & 1.257(7) & 0.39(2) & 0.27 & 0.575(4) & 0.22(1) & 0.15 & 0.253(3) & 0.11(1) & 0.07 \\
``Bulgy''           &  6680 & 2.68(1)   & 0.74(3) & 0.52 & 1.198(5)  & 0.48(1) & 0.34 & 0.547(3) & 0.28(1) & 0.20 & 0.238(3) & 0.15(1) & 0.11 \\
``Disky''           &  17596 & 2.030(8)   & 0.67(2) & 0.67 &  0.842(4)  & 0.45(1) & 0.45 & 0.349(3) & 0.28(1) & 0.28 & 0.146(2) & 0.14(1) & 0.14 \\
``Pure disk''    &  8168 & 1.91(1)   & 0.57(2) & 0.57 &  0.766(7)  & 0.40(1) & 0.40 & 0.303(4) & 0.26(1) & 0.26 & 0.121(4) & 0.13(1) & 0.12  \\
\hline
\end{tabular}
\end{table*}

Figure \ref{c1} shows that as we move from late--type spirals (with small bulges) to early-type spirals (with large bulges) the average face-on colours ($C_\circ$ in Table \ref{ctable}) reddens significantly. In fact {\it the average face-on colours of the ``very bulgy" subset are redder than the average edge-on colours of the ``pure disk" subset} (by 0.3, 0.1, 0.01 and 0.002 mags in $(u-z)$, $(g-z)$, $(r-z)$ and $(i-z)$ respectively) - although the spread of the colours of even face-on ``disky" spirals is large enough to almost encompass the spread of face-on ``bulgy" spiral colours. As discussed above, dust reddening appears again to be about equal in predicting the colors of galaxies as the bulge size and the stellar mix of the galaxy (however the difference is its systematic trend with viewing angle). 

We observe that the slope of the reddening with axial ratio is steepest for the ``bulgy" subsample. If we assume that a ``disky'' spiral (\ie~ one with $f_{DeV} < 0.5$) has an intrinsic axial ratio of $q\sim0.1$, 
while a ``bulgy'' spiral has $q\sim0.2$ (see Table 1 for a justification of these choices), 
then the total reddening from face-on to edge-on (noted as $\Delta_0$ in Table \ref{ctable}) is given by $\gamma$ for the ``disky'' (late-type) spirals and by $0.7\gamma$ for the ``bulgy'' (early-type) spirals. Using this assumption, we argue that the total reddening from face-on to edge-on is always larger in the ``disky'' spirals than in the ``bulgy'' ones which then indicates a larger role for dust reddening on the average colours of the ``disky" spirals than in the ``bulgy" spirals. 

We test this data against the photometric models of \citet{T04} in Section 4 below.

\subsection{Dependence on Galaxy Luminosity}

Luminosity dependence of the amount of dust attenuation in spiral galaxies was studied in NIR bands in \citet{G95} and \citet{M03} and in the optical/NIR in \citet{T98}. \citet{M09} have also studied the impact of luminosity and face-on colour on reddening trends of late-types in SDSS bands. In the previous section we showed differences in the trends of reddening with inclination for GZ spirals with different bulges sizes (as measured with $f_{DeV}$ from SDSS). It has been shown many times that different types of spirals have different luminosity functions (\eg ~work using 1500 SDSS galaxies with visual classification, \citealt{N03}), so some, or all of this trend could be due to the different luminosity ranges of spirals with different bulge sizes.

\begin{figure*}
\includegraphics{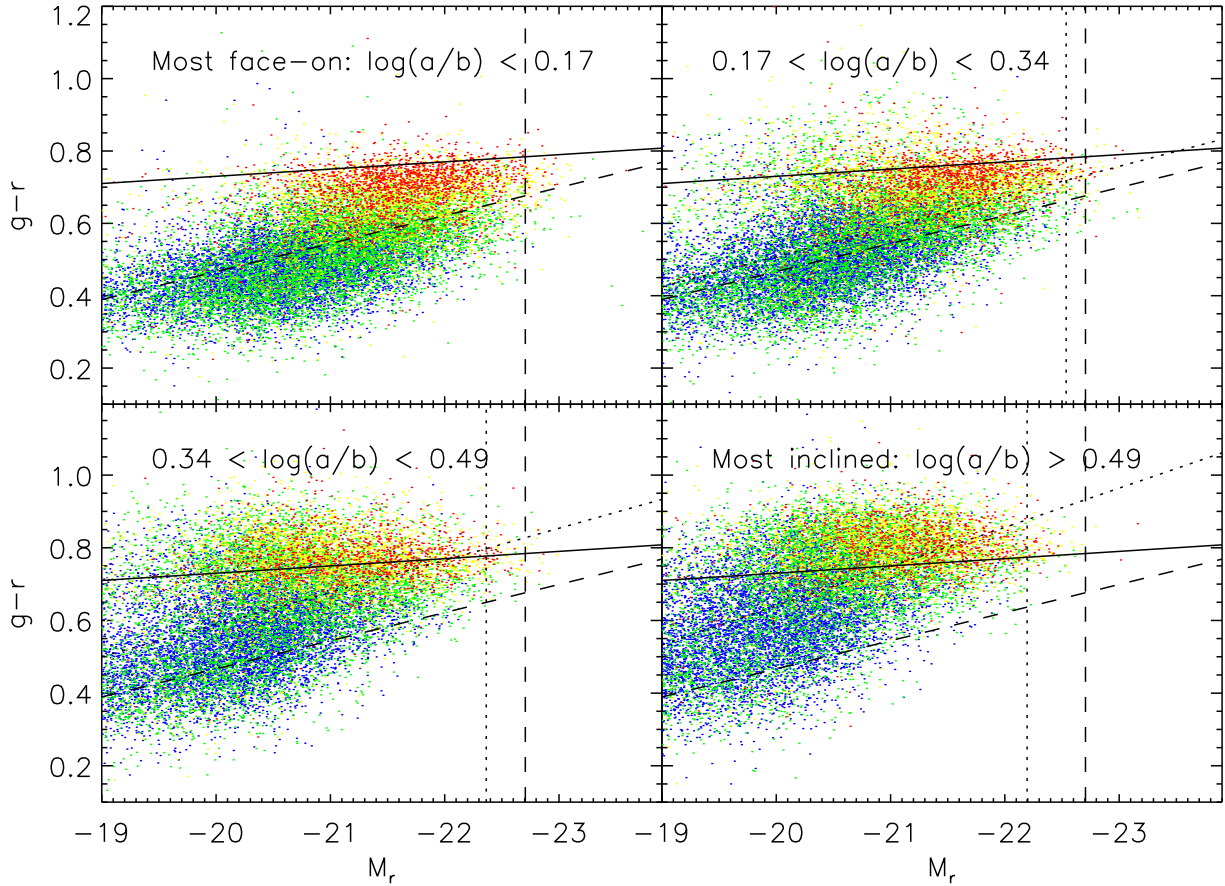}
\caption{Colour-magnitude ($M_r$ vs. $(g-r)$) for Galaxy Zoo spirals in four quartile bins of axial ratio (as indicated). Each plot has $\sim 20000$ galaxies on it. Neither colour nor magnitude is corrected for internal extinction. The galaxies are colour coded by $f_{DeV}$ using: red: $f_{DeV}>0.9$; yellow: $0.5< f_{DeV}>0.9$; green: $0.1< f_{DeV}>0.5$; and blue: $f_{DeV}<0.1$. The solid line indicates the blue edge of the red sequence of GZ ellipticals. The dashed line in each panel is a fit to the blue cloud in the most face-on quarter of Galaxy Zoo spirals ($\log(a/b)<0.14$) - with the vertical dashed line indicating the 99.5th percentile magnitude (\ie only 5\% of GZ spirals in this axial ratio bin have $r$-band magnitudes brighter than that. The dotted lines in the 3 panels showing more inclined spirals are the same quantities for those galaxies and illustrate the change in the shape of the CM diagram of spirals as they become more inclined.
\label{colourmag}}
\end{figure*}

 In Figure \ref{colourmag} we show the observed (\ie~ not corrected for internal extinction) colour magnitude (CM) diagram of all GZ spirals ($p_{\rm spiral}>0.8$, $0.01<z<0.09$) in four quartile bins of axial ratio. Each plot has $\sim 20000$ galaxies on it. The galaxies are colour coded by $f_{DeV}$ using: red: $f_{DeV}>0.9$; yellow: $0.5< f_{DeV}>0.9$; green: $0.1< f_{DeV}>0.5$; and blue: $f_{DeV}<0.1$. The solid line indicates the blue edge of the red sequence of GZ early-types at $z=0$.
 The ``face-on" GZ spirals are mostly to the blue side of this line, but in the most edge-on quartile many galaxies are reddened across it. We note that most spirals whose intrinsic colours place them in the so-called ``green valley" are bulge-dominated, but there is significant contamination to this part of the CM diagram by dust reddened inclined spirals. 
 
 The dashed line in each panel of Figure \ref{colourmag} is a fit to the blue cloud in the face-on GZ spirals ($\log(a/b)<0.14$) - with the vertical dashed line indicating the 99.5th percentile magnitude (\ie~ only 5\% of spirals have $r$-band magnitudes brighter than that line). The dotted lines in the 3 panels showing more inclined spirals are the same quantities for those inclined galaxies and illustrate the change in the shape of the CM diagram of spirals as they become more inclined. In the most inclined quartile the 95th percentile $r$-band magnitude is 0.5 mag fainter than in the face-on spirals, quite consistent with the estimate of the trend of $r$-band magnitude with axial ratio we derived in Section 3.1 above.
 
\begin{figure}
\includegraphics[width=84mm]{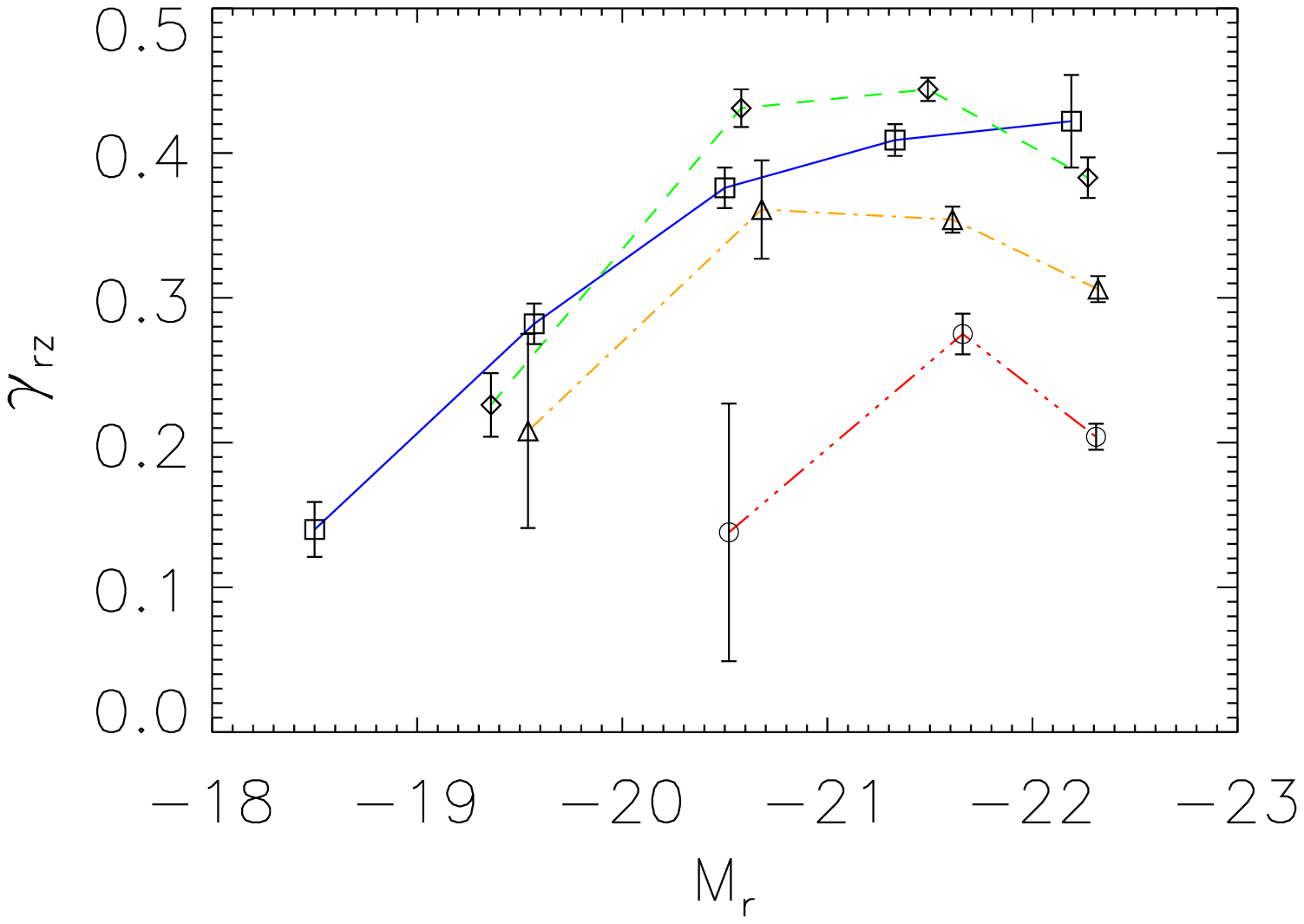}
\caption{Slope of the trend of $(r-z)$ reddening with axial ratio plotted against the $r$-band absolute magnitude (with a basic correction for dust as described in the text). The different lines show subsets divided by $f_{DeV}$ as used elsewhere in this work going from blue--green--orange--red (solid--dashed--dot-dash--dot-dot-dot-dashed; symbols square-diamond-triangle-circle) from most disk dominated to most bulge dominated. \label{gamma}}
\end{figure}

We derive the face-on $r$-band absolute magnitudes of the well resolved GZ spirals using a correction of $\Delta M_r = 0.8 \log(a/b)$ to $\log(a/b)=0.7$, and $\Delta M = 0.56$ after (as suggested in Section 3) then fit linear relations to the colour versus axial ratio in subset of the GZ spirals separated by both bulge size (as measured by $f_{DeV}$) and absolute magnitude. We plot the slopes of these relations for the $(r-z)$ colour versus $r$-band magnitude in Figure \ref{gamma}. Interestingly the slope of the trend of reddening with axial ratio shows hints of a peak in all spiral type subsets at around $M_r \sim -21.5$ where it is roughly twice as large as at the lowest luminosities. At both dimmer and brighter magnitudes the reddening trend seems to decrease (note that in the subset of the most bulge dominated spirals there are very few galaxies with $M_r > -21$). We comment that while we have selected the galaxies in these magnitude bins after an estimated dust correction has been applied, the implicit magnitude limit of the original selection may be causing some bias here (i.e. causing edge-on galaxies which would make it into the sample if seen face-on to be dimmed below the selection limit) and would act in a way to cause an underestimate of the slope of reddening (since brighter galaxies will on average be redder). However the blue cloud shows only a mild trend of colour with luminosity, so we argue that this effect is unlikely to account for all of the trend with luminosity we observe here. However the bias may be larger at the faint end of the sample, so the downturn in reddening at in low luminosity spirals should be interpreted carefully.

At a given absolute magnitude the slope of the reddening is significantly shallower (0.1-0.2 mag/dex) in the ``very bulgy'' subset $f_{DeV}>0.9$ than in more disk dominated spirals. This means that a single correction to magnitudes and colours for dust will be an overestimate of the effect in both very dim and very bright (especially bulge dominated) spirals and an underestimate especially for disk dominated spirals with $M_r \sim -21.5$. Interpreting these slopes in terms of dust content is not trivial as illustrated in Section 4. The slope of the predicted trend of $(r-z)$ colour from the models of \citet{T04} is degenerate with bulge-disk ratio and face-on central opacity. However in pure disk galaxies which our $f_{DeV}<0.1$ sample should approximate (shown in blue in Figure \ref{gamma}) an increase in the slope of the trend of reddening can be interpreted as an increase in dust content in the galaxies. 

 An increase in dust content is expected for more luminous galaxies which have had more star formation over cosmic time and which also are physically larger and so their path lengths increase. A decrease of the trend for the most luminous galaxies has not been observed before, perhaps because previous studies have focussed on later type spirals which are dominated by the lower luminosities. However it has been observed that lower luminosity galaxies have more active {\it recent} star formation than high mass galaxies even amongst spirals (eg. \citealt{Y99}). Dust is destroyed over time \citep{D79}, so these hints in the downturn in the slope of reddening in the most luminous galaxies may be related to their lower levels of {\it recent} star formation. 

 Clearly the total reddening of a spiral galaxy with inclination is a complex mixture of it's dust content (via its star formation history), physical size and dust geometry, all of which may depend on other global quantities such as luminosity, or morphological type. In order to understand dust attenuation we therefore argue a two pronged approach is needed. Individual modeling of galaxies using multi-wavelength data (from UV to far-IR) is needed to shed light on the details of the dependence of dust on galaxy properties; while the kind of statistical approach used in this paper is needed to look at global trends in large samples of galaxies, and to provide empirical corrections of use to remove, at least to first order, the bias introduced into galaxy surveys from the inclination dependent attenuation and reddening of spiral galaxies. 

\subsection{Galaxy Zoo Red Spirals \label{massredness}}

In \citet{B09}, we discuss the existence of a large population of ``red spirals" detected in the GZ sample, i.e., galaxies with colours consistent with elliptical galaxies ($(u-r)$ colour greater than the dividing line in Figure 7c of \citet{B06}) but with clear evidence of a disk and/or spiral arms. We revisit these ``red spirals" here mainly to note that the dividing line between ``red" and ``blue" changes as a function of the derived stellar mass of the galaxy, the measurement of which can depend on spiral galaxy inclination. The presence of dust both increases the mass-light ratio (redder colours imply higher $M/L$) and decreases the luminosity (due to extinction) so a trade off exists as discussed by other authors \citep[e.g.][]{D08,M09}. In \citet{Sk09}, we presented the clustering of the red spirals (now selected from the $M_r$ versus $(g-r)$ colour-magnitude and examined the expected contamination of this sample by reddened edge--on spirals. 
  
We emphasize that we still find a population of interesting ``red spirals". For example, we see face-on GZ spirals that are still redder than the normal colour dividing line between spirals and ellipticals (eg. Figure \ref{colourmag}), and even the reddest edge-on spirals must be unusually passive unless they have significant amounts of dust. Furthermore, \citet{B09} demonstrated that the face-on red spirals have similar environmental dependencies as all red spirals (including the dust reddened spirals). \citet{redspirals} study in more details these intrinsically red spirals.

\section{Photometric Models for Dust Attenuation in Spiral Galaxies}

We next compare models for dust attenuation with our observed relations. There has been significant progress in modeling the attenuation of the total light from spiral galaxies as a function of inclination, and models continue to grow in complexity. 
Both \citet{F99} and Tuffs et al. 2004 (hereafter T04) provide models which allow the attenuation to be calculated for a spiral galaxy at any viewing angle and with a given bulge-to-disk ratio. 
For a history of dust modeling in spiral galaxies see the introduction of either T04 or \citet{P04}. A more recent discussion of the T04 model can be found in \citet{PT07}.

Here we compare our data to the T04 model. In this model, a galaxy can be constructed with any bulge-disk ratio. The attenuation of each of three components (a bulge, disk and thin disk) is provided as a functional fit with inclination at central optical depths in B-band of $\tau_B = 0.1$, 0.3, 0.5, 1.0, 2.0, 4.0 or 8.0. The different components are then combined in the appropriate fractions to give the total attenuation. The output of the T04 model is published in various UV bands, as well as BVIJK. We linearly interpolate these results to the central wavelengths of the SDSS $ugriz$ bands (as recommended by Tuffs, 2009, priv. comm.). 

When attenuation curves are considered (\ie~ attenuation as a function of inclination) the relative geometry of the dust and stars are just as important as the wavelength dependence of the extinction (which is ``Milky Way like'' in T04).
Details of the exact distributions and relative sizes of the diffuse stellar light and dust can be found in T04. 
For our application it is important to know that the relative geometry is fixed based on optical/NIR imaging of a small sample of edge-on galaxies \citep{X99}, and that also the relative face-on B-band optical depth of the thick and thin disks is fixed at a ratio of 0.387 as measured for the edge-on galaxy NGC 891 \citep{P00}. 

The free parameters are then the total face-on B-band opacity ($\tau = \tau_{\rm disk} + \tau_{\rm thin disk}$), how ``clumpy" the clumpy dust component is (expressed by the clumpiness factor, $F$, for which a suggested value of $F=0.22$ is given in T04 based on observations of NGC 891), and the inclination of the galaxy. The observed attenuation also depends on the observed bulge-disk ratio of the galaxy. 

One final ingredient which will be important in comparing the T04 model to our data is the dependence of intrinsic axial ratio on the intrinsic bulge-disk ratio. Only in the case of an infinitely thin disk is the inclination related to the axial ratio by $\cos(i) = b/a$. In normal disks with an intrinsic axial ratio, $q$, the inclination is given by Equation \ref{inclination}, and we expect that $q$ will increase monotonically with increasing bulge-disk ratio. 

\subsection{Direct Comparison}
\begin{figure}
\includegraphics[width=84mm]{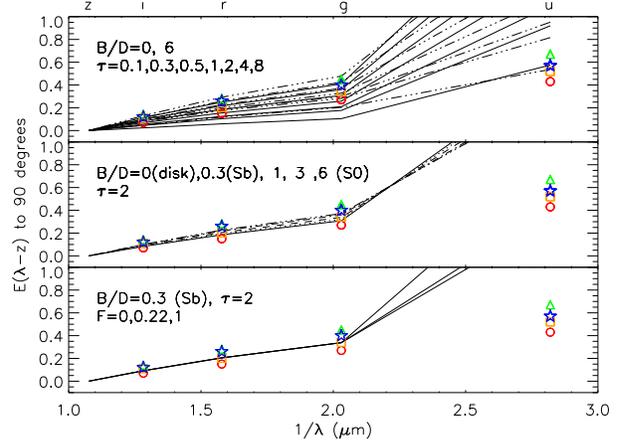}
\caption{The average reddening $\lambda-z$ from face-on ($i=0\deg$) to edge-on ($i=90\deg$) from the models of T04. {\it Top panel:} the results for a pure disk model ($B/D=0$, solid line) and a large bulge-disk ratio ($B/D=6$; dot-dot-dot-dashed line) for central B-band face-on opacities of 0.1, 0.3, 0.5, 1, 2, 4 and 8 (from lower to upper). {\it Middle Panel:} the results for varying the bulge-disk ratio from pure disk ($B/D=0$) to large bulge-disk ratios (up to $B/D=6$) when $\tau_B=2$. {\it Lower Panel: } the result of changing the clumpiness factor, $F$ with $\tau_B=2$ and $B/T=0.3$ (Sb like spirals). 
Overlaid with symbols are the results of our fits to the ``very bulgy'', ``bulgy'', ``disky'' and ``pure disk" subsets of the GZ spirals (circles, squares, triangles and stars respectively; see Table \ref{ctable}).
\label{T04model}}
\end{figure}

 In Figure \ref{T04model} we plot lines showing the dependence of the total reddening from edge-on to face-on with wavelength for the free parameters in the T04 model. The top panel is an illustration of the impact of changing the central B-band face-on opacity in the case of a pure disk model (solid line) and a galaxy with a large bulge-disk ratio ($B/D=6$, dot-dash line). In the middle panel the impact of changing the bulge-disk ratio from a pure disk galaxy through to large bulge-disk ratios for a model with $\tau^0_B = 2$ is shown. In the bottom panel we show the impact on $(u-z)$ reddening of the full range of possible values for the clumpiness factor F. This factor only affects the $u$-band in the SDSS filter.

Also plotted in Figure \ref{T04model} are fits to the observed total attenuations from Section \ref{3}. While the model can easily predict the attenuation in the optical colours, it is clear that the edge-on attenuation curve significantly over-predicts the observed edge-on $u$-band attenuation for reasonable values of the central B-band opacity. The $u$-band is the only SDSS band in which the thin disk contributes to the attenuation, which causes the attenuation to increase significantly only at inclinations where the thin disk is close to being viewed edge on. The exact wavelength transition between where the thin and thick disks are important is not well known. The T04 model can be modified to make the $u$-band be dominated by the thick disk (by multiplying the B-band output of the model by 1.267) and this provides a better fit to the GZ spirals (Tuffs \& Popescu, 2009, priv. comm.). 

\begin{figure}
\includegraphics[width=84mm]{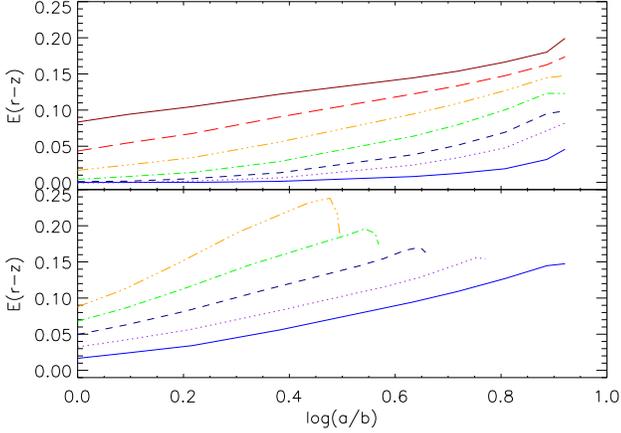}
\caption{The $(r-z)$ reddening as a function of observed axial ratio from the models of T04. {\it Top panel:} the results for a pure disk model ($B/D=0$) for central B-band face-on opacities of 0.1, 0.3, 0.5, 1, 2, 4 and 8 (from lower to upper). {\it Lower panel:} the results for $\tau_B=2$ and $B/T$ values of 0, 0.25, 0.5, 0.75 and 1.0. \label{T04modelrz}}
\end{figure}

 We show Figure \ref{T04modelrz} as an example of the effect of varying model parameters on the curves of optical reddening with axial ratio. We show this for a range of central B-band opacities (top panel) and for a range of bulge-disk ratios at $\tau_B=2$ (lower panel).  It is interesting to note that while increasing the central opacity (or the total amount of dust) increases the total amount of attenuation significantly, the slope of the curve of attenuation with axial ratio does not change much. Unfortunately this is the main observable we are sensitive to (not knowing {\it a priori}  the expected intrinsic colours of spiral galaxies), which (if the models are correct) will limit the information on $\tau_B$ which can be obtained from studying the inclination dependent optical attenuation of spiral galaxies (at least via colours). The effect of an increasing bulge-disk ratio increases the slope of the curve since at larger $B/D$ more inclined galaxies have smaller axial ratios. Also striking is the downturn in attenuation at large inclinations when bulges are present (as discussed in T04). 
 
The amount of face-on reddening from the T04 models is only a fraction of a magnitude even for $\tau_B=8$, showing that (even given the large trends of observed colour with axial ratio), dust reddening of face-on galaxies cannot be significant. The range in face-on colours predicted by T04 from central opacities varying from $\tau_B=0$-8 is also much smaller than the observed range of face-on colours of galaxies (\eg ~Figure \ref{colour_fdeV}). This illustrates that (if this model is correct) the colour of a face-on spiral galaxy is set by the intrinsic colour of its stellar population and not by the reddening effects of dust. This therefore shows (again if the T04 model is correct) that the face-on red spirals observed in Galaxy Zoo \citep{redspirals} must have intrinsically red colours from old stellar populations. 
 
\subsection{Monte Carlo Realisations}

To produce a fair comparison of our data with the models, we want to fully sample the intrinsic scatter in the properties of our GZ spirals. Therefore, we produce Monte Carlo realisations of the model attenuation curves in the SDSS bands based on reasonable assumptions about the galaxies. We use the correlations found in Section \ref{2.4} between $f_{DeV}$, bulge-disk ratio, intrinsic axial ratio and face-on colour to construct Monte Carlo estimates of the distribution of observed colours from the T04 models.

The Monte Carlo ingredients are:
\begin{enumerate}
\item[1. $\tau_B$] - assumed to be Gaussian with a mean tailored to fit the data and standard deviation of 1.
\item[2. F] - has no affect on attenuation curves, so set to F=0.22 
\item[3. $B/T$] - modelled using the observed distribution of $f_{DeV}$ and the relation between $f_{DeV}$ and $B/T$ we find in Section \ref{2.4} (adding in a scatter of 0.1 in $B/T$). 
\item[4. $\log(a/b)$] - from Eqn \ref{inclination}, assuming random inclinations and intrinsic axial ratios which depend on $f_{DeV}$ as found in  Section \ref{2.4}.2
\item[5. Face-on colours] - input as a function of $f_{DeV}$ based on observed colours of face-on GZ spirals (Figure 3).
\end{enumerate}
    
  We try various input values for $\tau_B$. For low values of $\tau_B$ (less than $\tau_B=1.5\pm 1$) the model is able to predict the observed trend of $(u-z)$ colours, but significantly under-predicts the optical colours of the most inclined galaxies. For any value of $\tau_B$ larger than about 1.5, the $(u-z)$ colour of the most inclined galaxies is over-predicted while the optical colours of the most inclined galaxies are under-predicted. (The over-prediction in $u$-band may be improved using the thick disk modification to the B-band attenuation in the T04 model which is discussed above.)

We do not reproduce the shape of the scatter in Figure 2, which shows hints of a bimodal pattern, and has a long trail to the blue side. The scatter in the Monte Carlo is dominated by our input assumption of a Gaussian distribution of face-on colours, so this shows a more complicated distribution of face-on colours is needed. The interquartile ranges matches well at low inclinations (which is expected as this is an input based on the data) but at large inclinations - where the effect of the model start to dominate -  the interquartile range it is found to be too large in the $u$-band colour and too small in the optical colours. This suggests that our Monte Carlo using the T04 model predicts a larger range of UV reddening and a smaller range of optical reddening than is observed. 
 
\begin{figure*}
\includegraphics{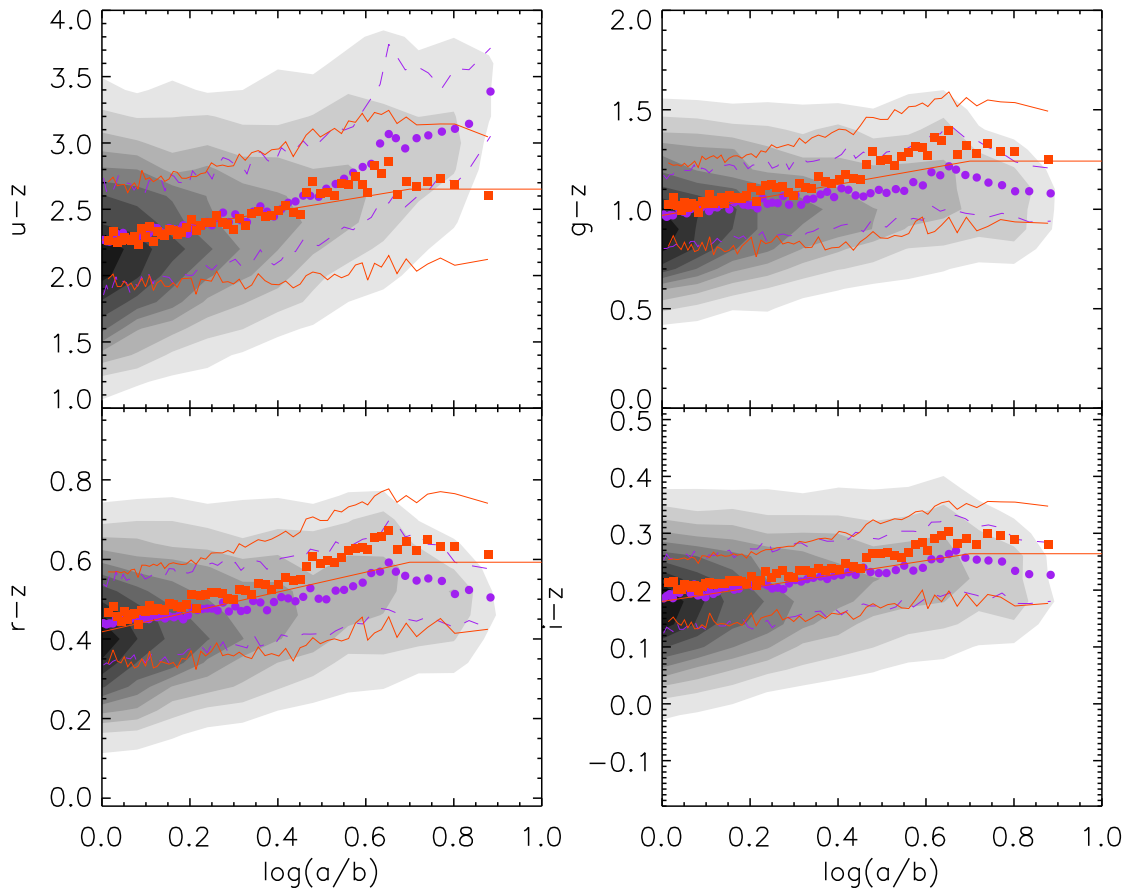}
\caption{Monte Carlo realisation of the reddening in SDSS colours of 24376 spiral galaxies based on the T04 model with $\tau_B = 4\pm1$. This figure should be compared to Figure \ref{colour} which shows the same thing for the real data. We over-plot in orange the median and interquartile range of the binned real data as well as a linear fit to the real data. In purple is shown the same thing for the Monte Carlo generated galaxies. 
\label{montecarlo}}
\end{figure*}

 Figure \ref{montecarlo} shows our model Monte Carlo realisations for the reddening as a function of  inclination with an input central face-on B-band opacity of $\tau_B = 4\pm1$. We pick this value of $\tau_B$ in part based on the $\tau_B = 3.8 \pm 0.7$ found by \citet{D07} using the same \citet{T04} models and looking at the inclination dependence of the B-band luminosity function, but it does also appear to describe the observed trends well. We note here that comparisons of our data with this model are therefore implicitly also comparisons between the the trends we present and those presented in \citet{D08} which are based on the T04 model with this value of $\tau_B$. We overplot (in orange) the binned real data and linear fit to that data. In purple, we show the same quantities but for the Monte Carlo realisations. 
 
 We split the sample by $f_{DeV}$ as done in Section 3.1 (see Figures \ref{c1} and \ref{c2}-\ref{c4} for the data). Here we just explore the two extreme subsets, the ``pure disks" ($f_{DeV} < 0.1$) and the ``very bulgy" ($f_{DeV}>0.9$) spirals (see Figures \ref{m1}  and \ref{m4}). This split removes worries about the conversion between inclination and observed axial ratio, as all galaxies in these two extremes of bulge size should have similar observed axial ratios at a given inclination. 
  
  In summary we find that the model reproduces the observed trends surprisingly well, considering the number of fixed geometrical parameters (eg. the relative size of the stellar and dust disks, bulge shape, etc.) and agrees with the value of $\tau_B = 3.8\pm0.7$ (central optical depth) found by \citet{D07} for the same models. However, the models overpredict the amount of $u$-band attenuation in edge-on galaxies while underpredicting in $gri$-bands for any reasonable choice of central opacity and bulge-disk ratio (but see previous comments about modifying the published model to better match our $u$-band data). This could point to an inadequacy of the Milky Way extinction law when applied to external galaxies, but more likely indicates that a wider range of model geometries need to be allowed for. The relative geometry of the dust and stars is as important as total dust content when considering the integrated colours of galaxies, and may differ quite substantially among spirals. For example  \citet{H09} study the dust distribution in a backlit galaxy observed with ACS, and find it has a very extended dust distribution, clearly illustrating the limitations of using a fixed geometry for all galaxies.  \citet{W92} show how very dusty galaxies can be quite blue if the stars extend well past the dust so that there is a blue contribution from the outer disk. The same effect might be observed if the dust is very patchy. The observed light will become progressively dominated by emission from stars in the clear regions as the wavelength shortens. This could easily explain the failure of the T04 model in the SDSS $u$-band. The increased scatter in the reddening might also be related to the patchy distribution of dust in spiral arms and the expected variation of the relative orientation of spiral arms with the major axis of the inclined spiral. 

\section{Summary and Conclusions}

We present here a careful study of 24,276 ``well-resolved" (significantly larger than the average SDSS seeing) spiral galaxies selected from the visual classifications of the Galaxy Zoo project. This sample avoids the effect of seeing on the axial ratios of these galaxies, thus allowing us to use them as a reliable proxy for inclination (see Figure 1). We then study the properties of these spirals as a function of inclination to infer the effects of galactic dust on these properties. We also compare the data with the latest models for dust attentuation. Our major results and conclusions are:

\begin{itemize}

\item{We find the SDSS parameter  $f_{DeV}$ (fraction of the light profile best fit by a de Vaucouleurs model) is a good candidate for estimating the bulge size (and thus spiral type) for our galaxies, and appears to be better than other morphological proxies commonly used in the literature (e.g. concentration has a significant dependence on axial ratio with more inclined spiral galaxies being more concentrated). By matching to a sample of MGC galaxies (with known B/T ratios) we find that $f_{DeV}$ correlates with the bulge size with a linear relationship of $f_{DeV} = 0.11(3) + 1.7(2)B/T$. As a confirmation of this result, we see that the face-on colours of GZ spirals show a clear trend with $f_{DeV}$ such that galaxies with larger $f_{DeV}$ are redder as expected for spirals with larger bulges (see Figure \ref{bulge}).}

\item{We observe a clear correlation between the colours of our spirals and the inclination of the disk, i.e., those with larger axial ratios (more inclined) are on average redder than those which are viewed face-on (with small axial ratios).  We have fit these relations and provide them in Section 3.1. We then use these fits to estimate  the total average extinction of spiral galaxies from face-on to edge-on, which is 0.7, 0.6, 0.5, 0.4 magnitudes in $ugri$ bands respectively (assuming 0.3 magnitudes of extinction in $z$-band).} 

\item{Using $f_{DeV}$, we have split the spiral sample into ``bulgy" and ``disky" spirals (or early- and late--type spirals). We see  that the average face-on colours of the ``bulgy" spirals is redder than the average edge-on colours of the ``disky" subsample (by as much at 0.3 mags in $(u-z)$). Likewise, the scatter in colors seen in Figures 3 and 6 is slightly greater than the trends with inclination or increased reddening. Therefore we show that the colours of a spiral galaxy are set about equally by the bulge-to-disk ratio, or spiral type (via the stellar populations in the disk and bulge) and reddening as a function of inclination, due to dust. However we comment that dust effects are systematic with inclination and therefore need to be accounted for to make unbiased samples.}  

\item{We explore the luminosity dependence of the reddening and show that the slope of the trend peaks in all types of spirals (as split by $f_{DeV}$) at around $M_r \sim -21.5$. An increase in the slope with luminosity has been observed before \citep{G95,T98,M03}.  An increase in dust content is expected for more luminous galaxies which have had more star formation over cosmic time and which also are physically larger and so their path lengths increase. A decrease of the trend for the most luminous has not been observed before, and is mostly likely caused by the lower levels of {\it recent} star formation which are observed in the most massive spirals (since dust is destroyed over time).}
 
\item{We compare our data with state-of-the-art dust attenuation models from T04. We construct a Monte Carlo realisation of the GZ spiral samples using reasonable estimates for the input range of inclinations (assumed to be random), bulge-disk ratios,  intrinsic axial ratios and face-on colours (taken from the observed distribution of $f_{DeV}$).  We find that the model reproduces the observed trends surprisingly well, considering the number of fixed geometrical parameters (eg. the relative size of the stellar and dust disks, bulge shape, etc.) and agrees with the value of $\tau_B = 3.8\pm0.7$ (central optical depth) found by \citet{D07,D08} for the same models. However, the models overpredict the amount of $u$-band attenuation in edge-on galaxies while underpredicting in $gri$-bands for any reasonable choice of central opacity and bulge-disk ratio. They also are not able to predict an increase in the range of the observed colours of galaxies with inclination. This could point to an inadequacy of the Milky Way extinction law when applied to external galaxies, but more likely indicates that a wider range of model geometries need to be allowed for. In fact modifying the T04 model to make the $u$-band emission come from the thick disk instead of the thin disk does appear to fit our data better (Tuffs \& Popescu, 2009, priv. comm.)} 
  \end{itemize}  

We end by emphazising the effect of inclination (and dust) on present and future large galaxy surveys.  As shown, the presence of dust in a galaxy both reddens and dims the emitted light, and this effect clearly increases for inclined spirals. This impacts any measurements based on the magnitudes and colors of galaxies, like stellar mass estimates, but more crucially can cause {\it systematic} incompletenesses in galaxy survey. In the era of high precision clustering measurements from galaxies (e.g. \citealt{P09}), such effects could become important to ensure a fair sample of galaxies is obtained, e.g., both volume and mass--limited samples will miss faint (low mass) inclined spirals which fall below the selection criteria because of dust extinction. This could induce small bias in the measurements of scale--dependent biasing of galaxies as the mass of spirals is correlated with environment (and preferentially in the field compared to ellipticals).

 \paragraph*{ACKNOWLEDGEMENTS.} 

 This publication has been made possible by the participation of more than 160,000 volunteers in the Galaxy Zoo project. Their contributions are individually acknowledged at \texttt{http://www.galaxyzoo.org/Volunteers.aspx}. Funding for the SDSS and SDSS-II has been provided by the Alfred P. Sloan Foundation, the Participating Institutions, the National Science Foundation, the U.S. Department of Energy, the National Aeronautics and Space Administration, the Japanese Monbukagakusho, the Max Planck Society, and the Higher Education Funding Council for England. The SDSS Web Site is http://www.sdss.org/. The Millennium Galaxy Catalogue consists of imaging data from the Isaac Newton Telescope
and spectroscopic data from the Anglo Australian Telescope, the
ANU 2.3m, the ESO New Technology Telescope, the Telescopio
Nazionale Galileo and the Gemini North Telescope. The survey
has been supported through grants from the Particle Physics and
Astronomy Research Council (UK) and the Australian Research
Council (AUS). The data and data products are publicly available from {\tt www.eso.org/$\sim$jliske/mgc/}. KLM acknowledges funding from the Peter and Patricia Gruber Foundation as the 2008 Peter and Patricia Gruber Foundation International Astronomical Union Fellow, and from the University of Portsmouth. MM and RCN acknowledge financial support from STFC. Support for the work of MM in Leiden was provided by an Initial Training Network ELIXIR (EarLy unIverse eXploration with nIRspec), grant agreement PITN-GA-2008-214227 (from the European Commission). CJL acknowledges support from The Leverhulme Trust and the STFC Science In Society Programme. Support for the work of KS was provided by NASA through Einstein Postdoctoral
Fellowship grant number PF9-00069 issued by the Chandra X-ray Observatory
Center, which is operated by the Smithsonian Astrophysical Observatory for and
on behalf of NASA under contract NAS8-03060.

We thank Richard Tuffs and Christina Popescu for providing extensive comments on this paper in advance of publication.

\appendix
\section{Concentration and Light Profile Shape versus Galaxy Zoo Morphologies \label{morphs}}

 Without access to Galaxy Zoo, or other visual classifications for the huge number of galaxies in SDSS it has become common practice to use concentration or other structural parameters (for example $f_{DeV}$ which describes the fraction of the light profile best fit by a de Vaucouleurs profile as opposed to an exponential profile) as a way  to separate early and late type galaxies. In this Appendix we show the relation between these two structural parameters and visual classifications from Galaxy Zoo as a guide to ease comparisons between samples split using concentration or $f_{DeV}$ and Galaxy Zoo morphologies.
 
  The first large comparison of $f_{DeV}$ and concentration versus morphology in SDSS galaxies can be found in \citet{S01}. There a sample of 287 visual classified galaxies and 500 spectrally classified galaxies were considered. \citet{Sh01} performed a similar study using a sample of 456 bright galaxies. Both of these papers prefer the use of concentration ($c=r_{90}/r_{50}$) as a morphological separator over $f_{DeV}$. \citet{S01} recommend $c\sim2.6$ as a separator, while \citet{Sh01} prefer $c\sim 3$. Both papers warn over the use of $f_{DeV}$ which in the brightest galaxies is dominated by the inner light profile shape. This means that large nearby spirals with bulges can have much larger values of $f_{DeV}$ than if the fit were averaged over the whole galaxy. Never-the-less $f_{DeV}$ has become a common parameter used as a proxy for morphology (\eg~ \citealt{UR08,M09}) perhaps because of the recognition that inclined spirals are more concentrated than those viewed face-on (\citealt{M09}, and see Section \ref{2.4} above). 
 
\begin{figure}
\includegraphics[width=84mm]{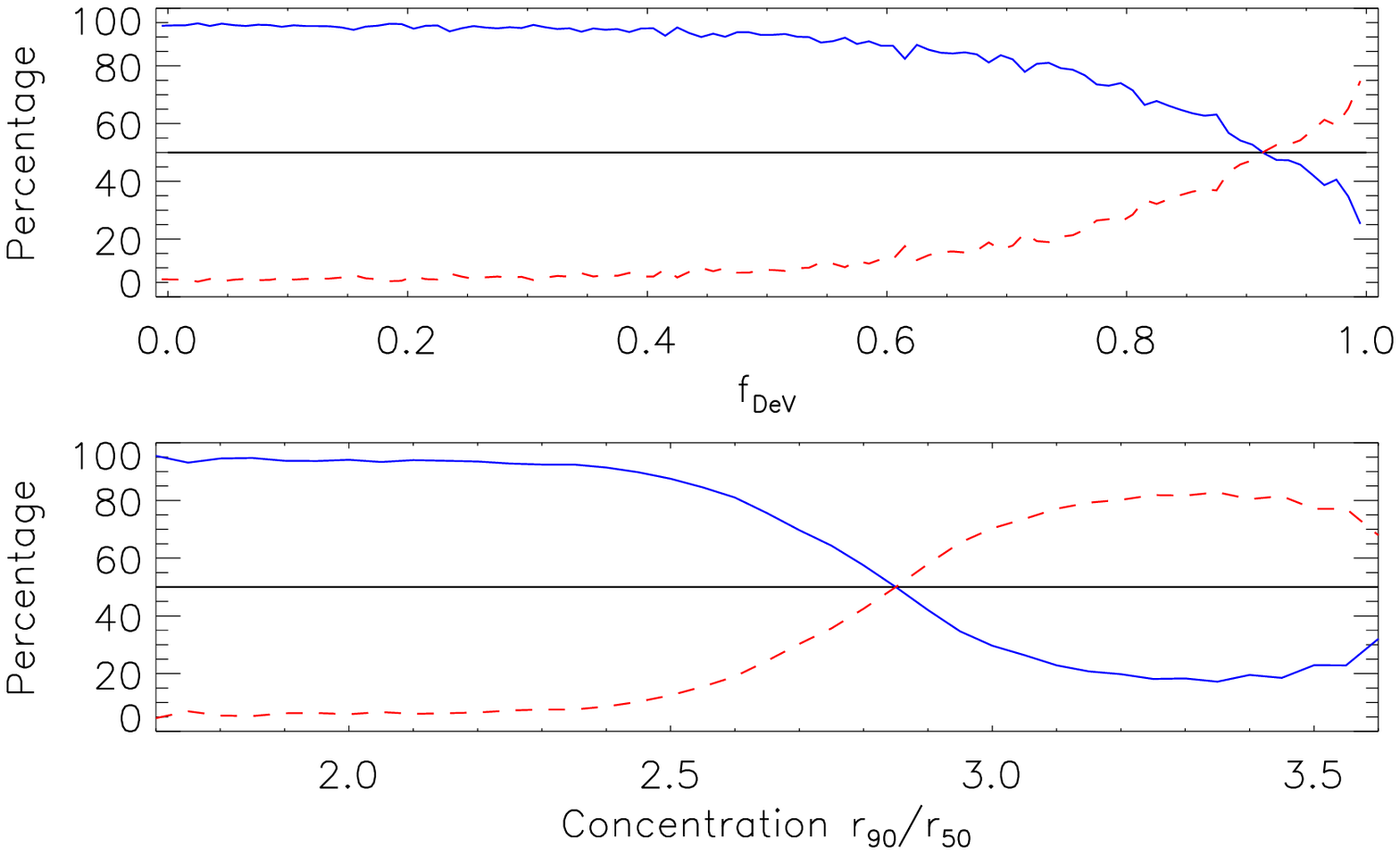}
\caption{The percentage of the ``clean" GZ sample (\ie~ that with either $p_{spiral}$ or $p_{ell} > 0.8$) falling into either the spiral (blue solid line) or elliptical (red dashed line) classification as a function of {\it Top: } $f_{DeV}$ (the fraction of the light profile fit with a de Vaucouleurs profile as opposed to an exponential disk) and {\it Bottom:} the concentration of the light. 
\label{frac_fdeV}}
\end{figure}

  Figure \ref{frac_fdeV} shows the fraction of the ``clean" GZ sample (\ie. those galaxies with very reliable spiral or elliptical classifcations; $p_{\rm spiral}>0.8$ or $p_{\rm ell}>0.8$) which are spirals (blue line) or ellipticals (red lines) as a function of $f_{DeV}$ or concentration. We see that while it is true that most ellipticals have $f_{DeV}\sim1$, many GZ spirals also have quite large values of $f_{DeV}$ (as is also evident in Figures \ref{bulge} and \ref{fdeV_fbulge} in Section 2.4 above). The lower panel shows the same fractions versus concentration and shows that the sample is dominated by spirals for $c<2.7-2.8$ and ellipticals at larger concentrations, but that there are still significant numbers of galaxies with very reliable spiral classifications from Galaxy Zoo (recall these all have $p_{\rm spiral} > 0.8$) with relatively large concentrations.
 
\begin{figure}
\includegraphics[width=84mm]{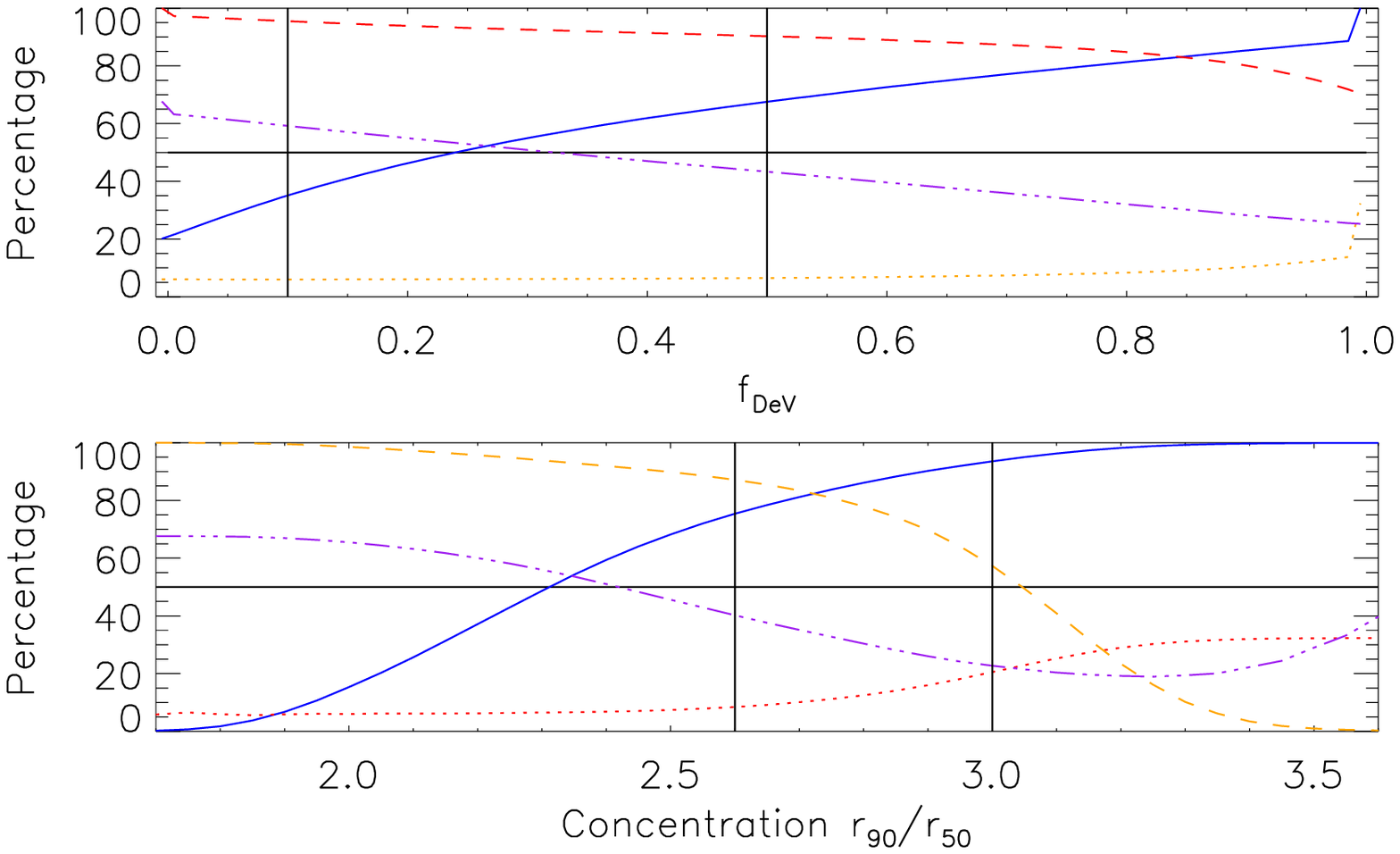}
\caption{Cumulative fractions for the fraction of GZ spirals (blue solid line) or ellipticals (red dashed line) found using limits on {\it Top:} $f_{DeV}$ or {\it Bottom: } concentration. For example, the upper plot shows that $f_{DeV}<0.5$ will find about 70\% of all the spirals, while $f_{DeV}>0.5$ selects about 90\% of the ellipticals. The orange dotted lines show the elliptical contamination to the ``spiral" samples, while the purple dot-dashed lines shows the spiral contamination to the ``elliptical" sample. For example the upper plot shows that for $f_{DeV}<0.5$, about 10\% of the selected galaxies are GZ ellipticals, and for $f_{DeV}>0.5$ about 40\% of the selected sample are GZ spirals. 
\label{cumulative}}
\end{figure}

 Figure \ref{cumulative} shows the cumulative fractions of samples of early types or spirals selected by limits on $f_{DeV}$ or concentration. Again red and blue lines show the fractions for the early types and spirals respectively. For example for an $f_{DeV}$ separator of 0.5 (as recommended by \citealt{S01}) this Figure shows that roughly 90\% of the early types are found, and 65\% of the spirals. The dotted lines show the percent early-type or spiral contamination in the samples. For this same example 45\% of the ``early types" found by $f_{DeV}>0.5$ are GZ spirals, while 5\% of the ``spirals" found by $f_{DeV}<0.5$ are GZ early types. In the terminology of \citet{S01} this shows the completeness and 100\% minus the reliability of the samples, and at given values of $f_{DeV}$ or concentration can be compared directly with the values in their Tables 2 and 3. 
 
  We find that $f_{DeV}$ is never a very reliable indicator of early types. The highest reliability which can be found is for $f_{DeV}\geq1$ in which roughly 70\% of the GZ early types are found, and the contamination by spiral types is 25\%. Furthermore GZ spiral galaxies with large values of $f_{DeV}$ are not always bright nearby galaxies as found by \citet{S01}. We find GZ spirals with large $f_{DeV}$ at the faintest magnitudes and largest redshifts in the GZ ``clean'' sample ($r\sim 17.7$ or $g\sim19$, and $z\sim 0.25$). On the other hand $f_{DeV}$ can be used fairly safely to identify spiral galaxies. Any cut on $f_{DeV}<0.6$ or so results in a sample which is 95\% GZ spirals and depending on the exact cut will have up to 70\% of the spirals. The most stringent cuts will miss large fractions of the spiral population, and (based on Figure \ref{fdeV_fbulge}) will preferentially select later and later type spirals (\ie ~with smaller bulges). For example \citet{UR08} use $f_{DeV}<0.1$ to select only the very latest type spirals, but we argue that even $f_{DeV}<0.5$ will miss the earliest spirals with the largest bulges.
  
  Selecting early types by concentration is also shown to be tricky. The $c>2.6$ suggested by \citet{S01} finds roughly 90\% of the GZ early types, but also results in a high contamination by GZ spirals (roughly 45\%). Many of these very concentrated spirals are likely to be edge-on and reddened, so an additional colour cut is unlikely to remove all of them without also a limit on the axial ratio (as also suggested by \citealt{M09}). Using $c>3$ as suggested by \citet{Sh01} finds 55\% of the early types, but results in a sample which is 25\% spiral. We find again that using concentration cuts to find spirals is more reliable, $c<3$ finds 95\% of the GZ spirals with 20\% contamination by early types, while $c<2.6$ finds 75\% of the spirals with a 10\% contamination. 
  
   With the large samples of galaxies now available (and planned in the future) there will always be a place for dividing galaxies in late/early subsets using structural parameters. In this Appendix we hope to have given greater insight into the levels of completeness and reliability which can be expected when using $f_{DeV}$ and concentration ($r_{90}/r_{50}$) from SDSS to do this. With Galaxy Zoo 2 data we will be able to further study this - looking at the types of spirals which end up contaminating the early-type subsets.
 
\section{Trends of SDSS colours: Extra Figures}

We show here the trends of the $u-z$, $r-z$ and $i-z$ colours with axial ratio where the GZ spirals have been split into rough spiral types using the SDSS parameter $f_{DeV}$ (Figs B1-B3). This is described in Section 3.1.1 of the text; the $g-z$ colour trends are shown in Figure \ref{c1}.

\begin{figure*}
\includegraphics{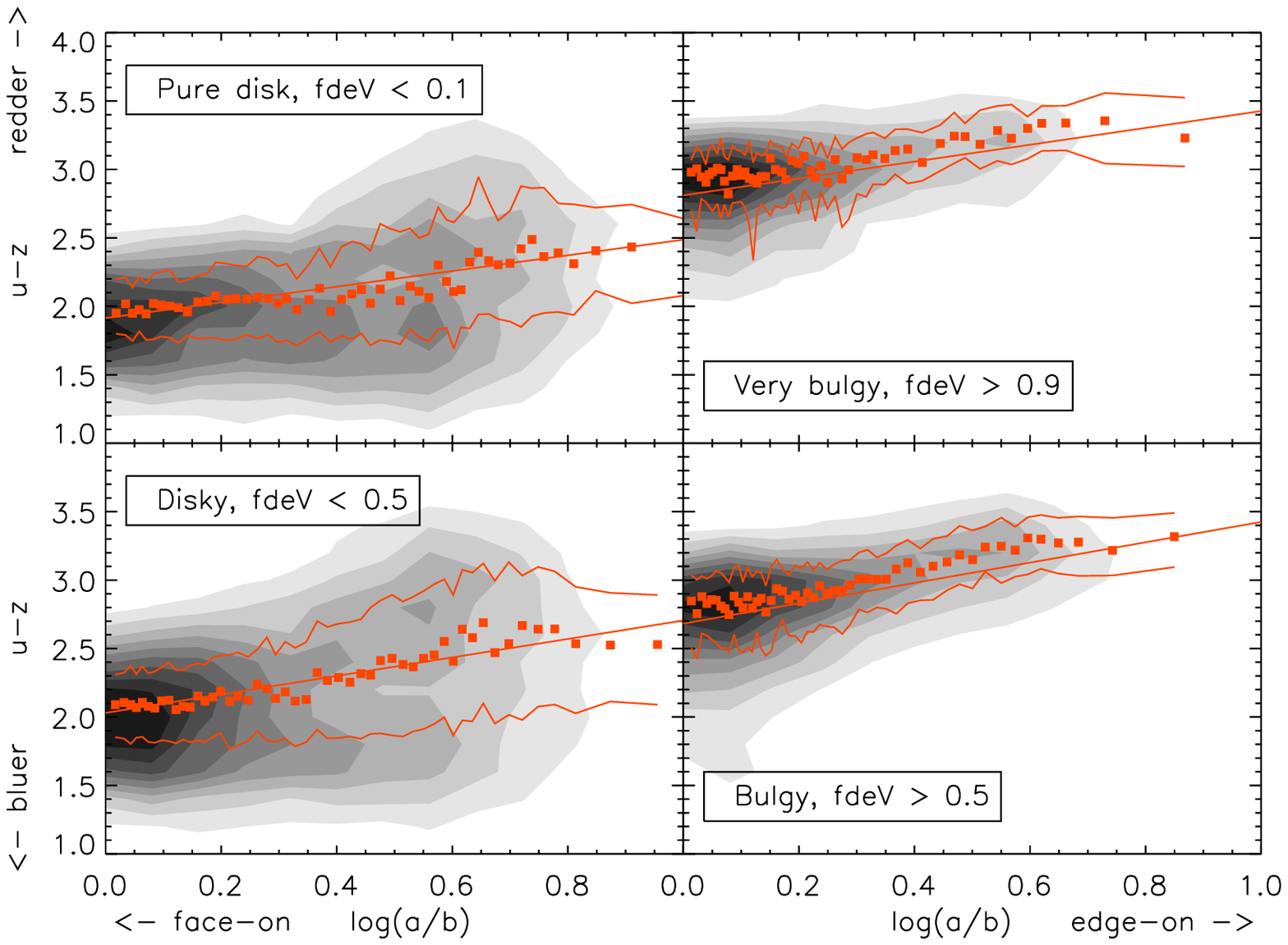}
\caption{The trend of $u-z$ colour for GZ spirals separated into rough spiral types using the $f_{DeV}$ parameter as indicated in the panels. As in Figure \ref{colour}, the greyscale contours indicate the locus of galaxies, while overlayed are the median and interquartile range of data binned in $\log(a/b)$.
\label{c2}}
\end{figure*}

\begin{figure*}
\includegraphics{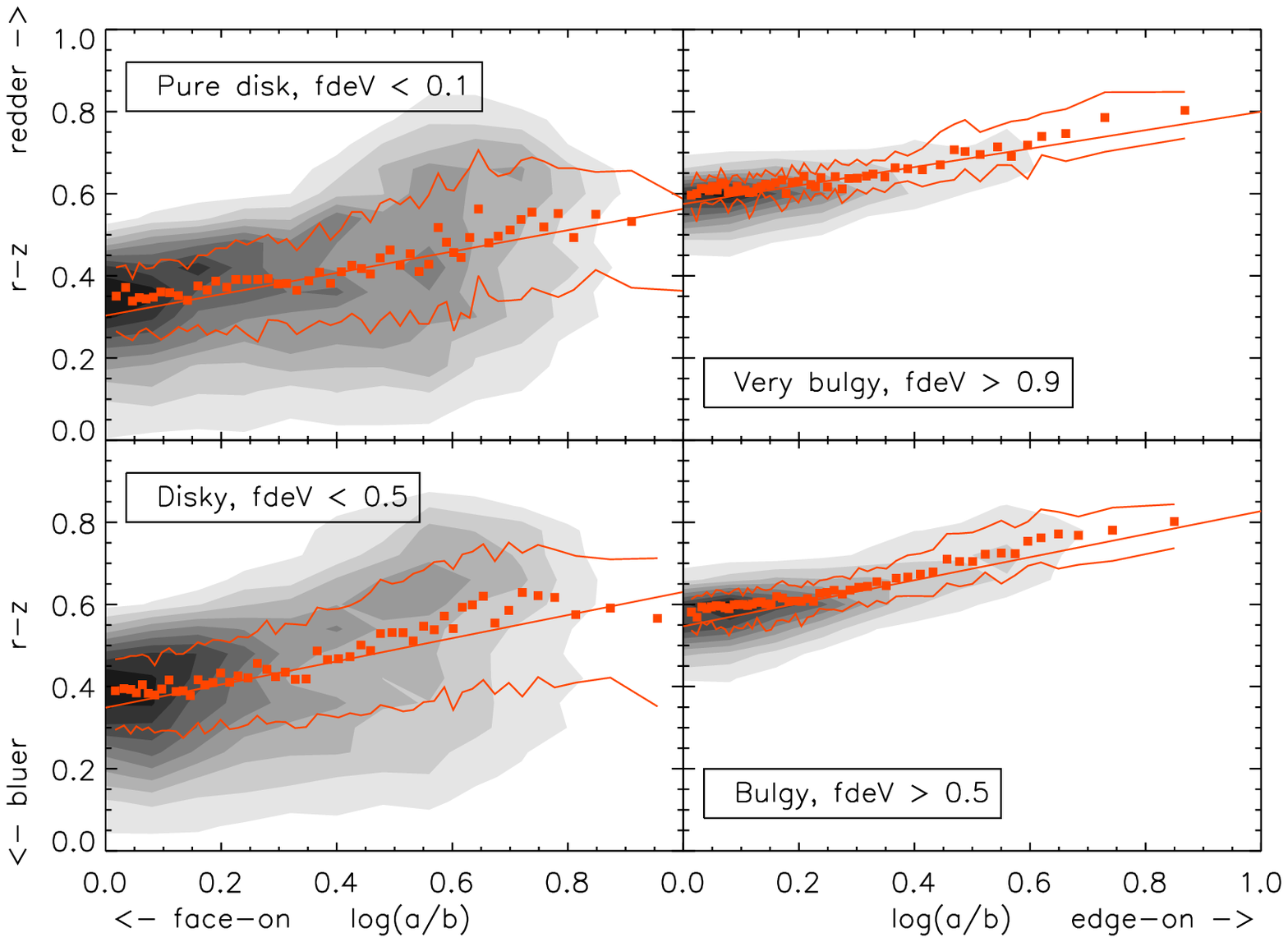}
\caption{The trend of $r-z$ colour for GZ spirals separated into rough spiral types using the $f_{DeV}$ parameter as indicated in the panels. As in Figure \ref{colour}, the greyscale contours indicate the locus of galaxies, while overlayed are the median and interquartile range of data binned in $\log(a/b)$.
\label{c3}}
\end{figure*}

\begin{figure*}
\includegraphics{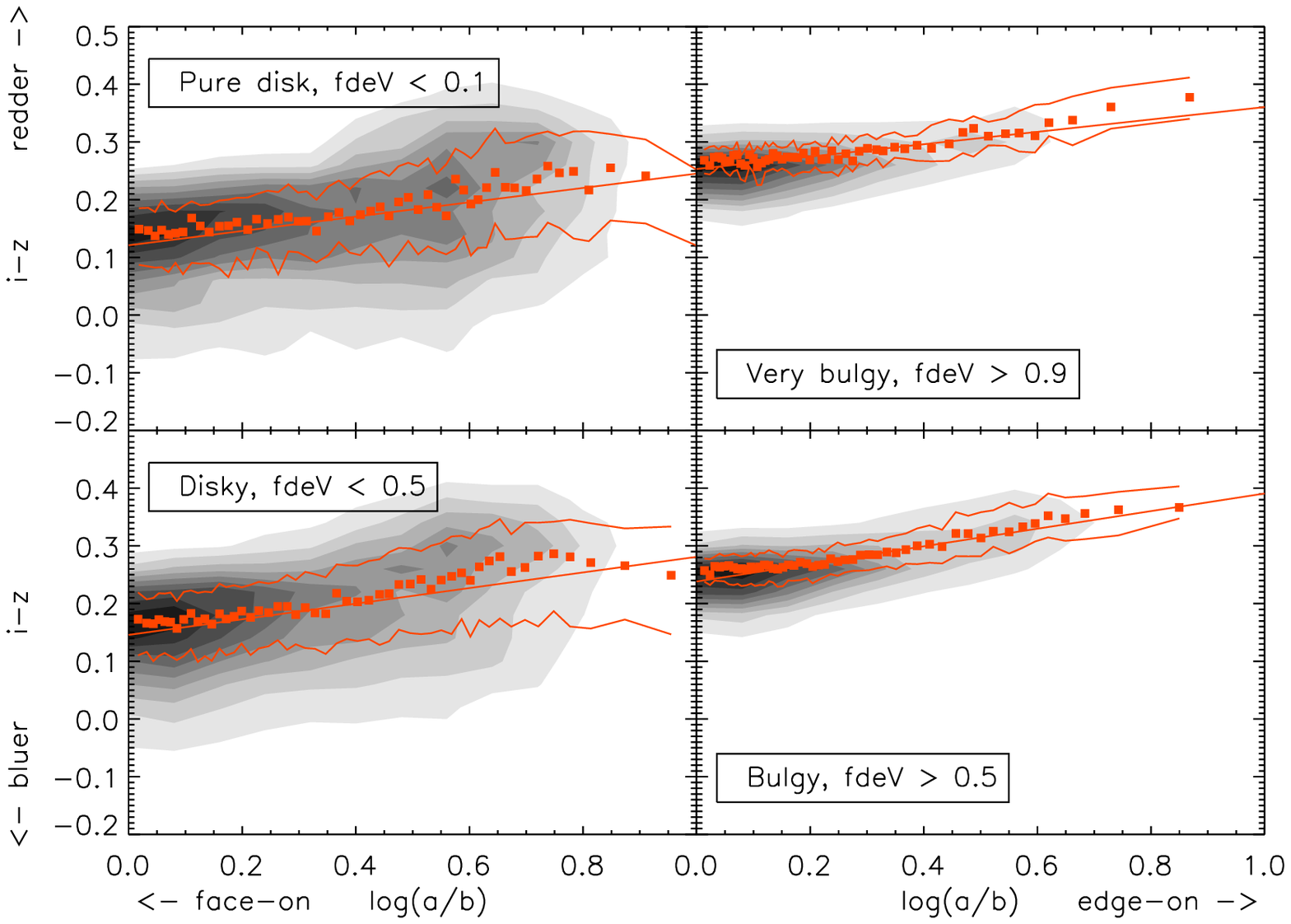}
\caption{The trend of $i-z$ colour for GZ spirals separated into rough spiral types using the $f_{DeV}$ parameter as indicated in the panels. As in Figure \ref{colour}, the greyscale contours indicate the locus of galaxies, while overlayed are the median and interquartile range of data binned in $\log(a/b)$.
\label{c4}}
\end{figure*}

\section{Monte Carlo Fits: Extra Figures}

\begin{figure*}
\includegraphics{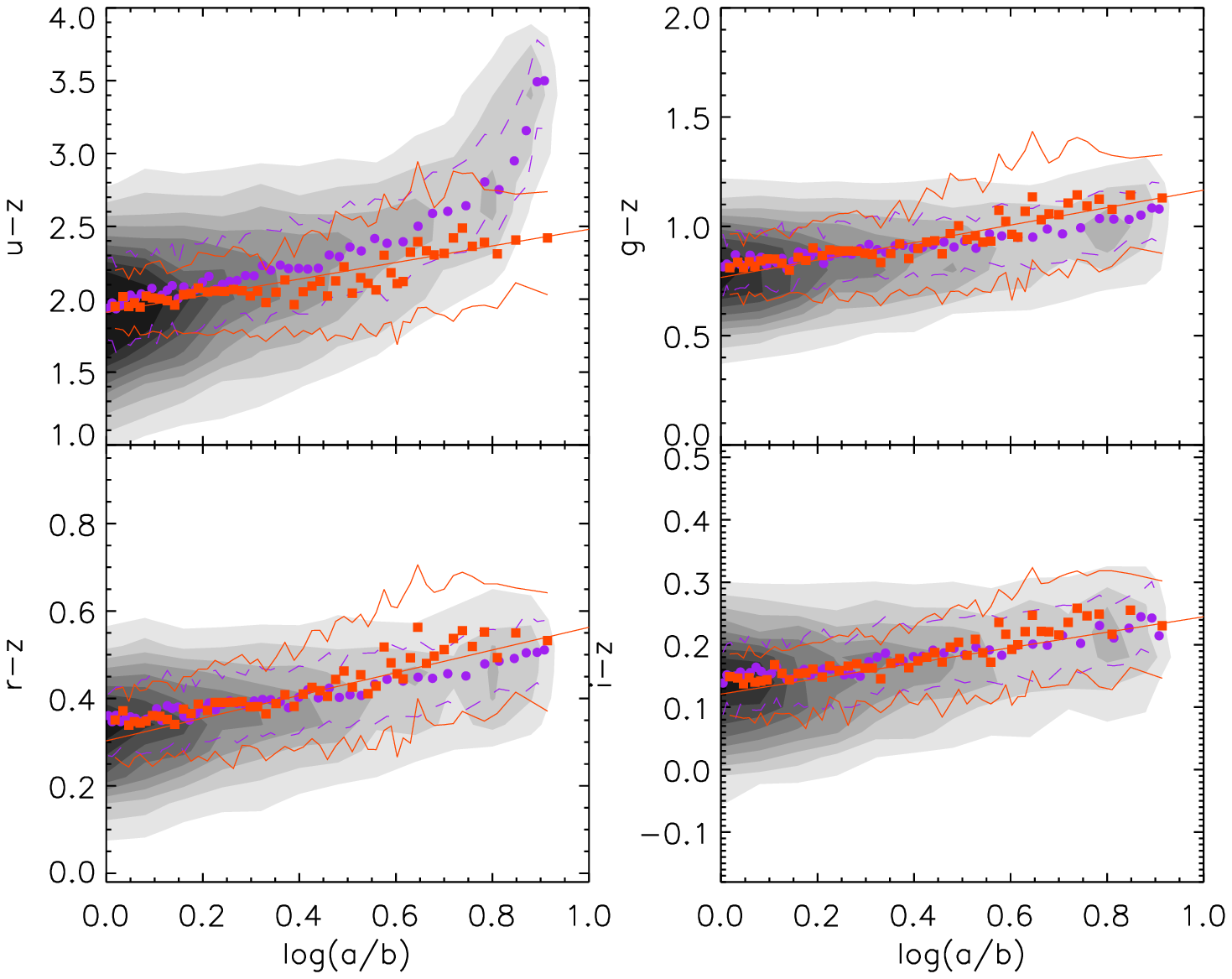}
\caption{As Figure \ref{montecarlo}, but only for galaxies with $f_{DeV} < 0.1$ - or "pure disk" spirals. Model parameters and Monte Carlo inputs are identical to that used for the full sample of GZ spirals.
\label{m1}}
\end{figure*}

\begin{figure*}
\includegraphics{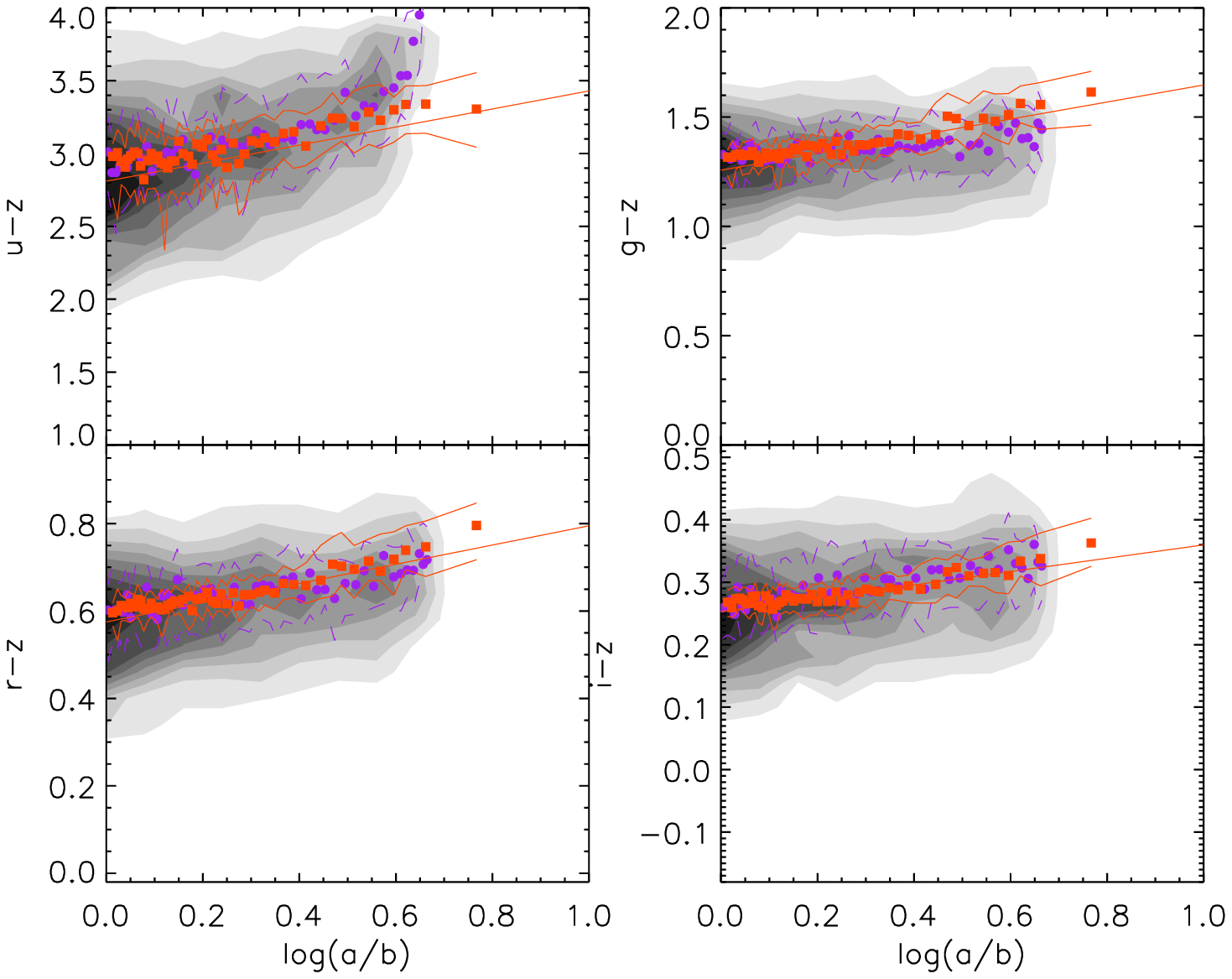}
\caption{As Figure \ref{montecarlo}, but only for galaxies with $f_{DeV} >0.9$ - or "very bulgy" spirals. Model parameters and Monte Carlo inputs are identical to that used for the full sample of GZ spirals.
\label{m4}}
\end{figure*}

 We show here in Figures \ref{m1}-\ref{m4} the results for Monte Carlo realisations of the data split into subsamples by $f_{DeV}$. Details are described in Section 4.1, and the results for the whole GZ spiral sample are shown in Figure \ref{montecarlo}.

\label{lastpage}

\end{document}